%% file: main.tex
\documentclass[journal]{IEEEtran}
\usepackage[latin1]{inputenc}
\usepackage{times,amsmath}
\usepackage{amssymb}
\usepackage{steinmetz}
\usepackage{pstool}
\usepackage{subfigure}
\usepackage{multirow}
\usepackage{enumerate}
\usepackage{graphicx}
\usepackage{multirow}
\usepackage{mwe}
\usepackage{bigints}
\usepackage{MnSymbol}
\usepackage{stfloats} 
\usepackage[table]{xcolor}
\usepackage[square, comma, sort&compress, numbers]{natbib}
\usepackage{nohyperref}
\usepackage{algorithm,algorithmic}
\usepackage{epstopdf}
\usepackage{mathtools, cuted}
\usepackage[table]{xcolor}

\graphicspath{ {Figures/} }

\begin{document}

\title{ You can monitor your hydration level using your smartphone camera }
%\title{ Non-Invasive Monitoring of Dehydration using a Smartphone Camera }
%\title{ Non-Invasive Assessment of Dehydration on a scale of 1 to 4 using an off-the-shelf Smartphone }

\author{
\IEEEauthorblockN{
Rose Alaslani\IEEEauthorrefmark{1}, Levina Perzhilla\IEEEauthorrefmark{1}, Muhammad Mahboob Ur Rahman\IEEEauthorrefmark{1}, Taous-Meriem  Laleg-Kirati\IEEEauthorrefmark{1},\IEEEauthorrefmark{2}, Tareq Y. Al-Naffouri\IEEEauthorrefmark{1} 
}

\IEEEauthorblockA{\IEEEauthorrefmark{1}Computer, Electrical and Mathematical Sciences and Engineering Division (CEMSE), \\ 
King Abdullah University of Science and Technology, Thuwal 23955, Saudi Arabia} \\
\IEEEauthorblockA{\IEEEauthorrefmark{2}The National Institute for Research in Digital Science and Technology, Paris-Saclay, France\\
\IEEEauthorrefmark{1}\{rose.alaslani,levina.perzhilla,muhammad.rahman,tareq.alnaffouri\}@kaust.edu.sa, }
\IEEEauthorrefmark{2}Taous-Meriem.Laleg@inria.fr

}

\maketitle

\input{abstract}
\input{sec1}

\input{sec3-1}

\input{sec4-1}

\input{sec4-2}

\input{conclusion}

\footnotesize{
\bibliographystyle{IEEEtran}
\bibliography{references}
}

\vfill\break

\end{document}

%% file: abstract.tex
\begin{abstract} 

%%% long abstract %%%

This work proposes for the first time to utilize the regular smartphone---a popular assistive gadget---to design a novel, non-invasive method for self-monitoring of one's hydration level on a scale of 1 to 4. The proposed method involves recording a small video of a fingertip using the smartphone camera. Subsequently, a photoplethysmography signal is extracted from the video data, capturing the fluctuations in peripheral blood volume as a reflection of a person's hydration level changes over time. To train and evaluate the artificial intelligence models, a custom multi-session labeled dataset was constructed by collecting video-photoplethysmogram data from 25 fasting subjects during the month of Ramadan in 2023. With this, we solve two distinct problems: 1) binary classification (whether a person is hydrated or not), 2) four-class classification (whether a person is fully hydrated, mildly dehydrated, moderately dehydrated, or extremely dehydrated). For both classification problems, we feed the pre-processed and augmented PPG data to a number of machine learning, deep learning and transformer models which models provide a very high accuracy, i.e., in the range of 95\% to 99\%. We also propose a highly-accurate alternate method where we feed high-dimensional photoplethysmography time-series data to a DL model for feature extraction, followed by t-SNE method for feature selection and dimensionality reduction, followed by a number of ML classifiers that do dehydration level classification. Finally, we interpret the decisions by the developed deep learning model under the SHAP-based explainable artificial intelligence framework. The proposed method allows rapid, do-it-yourself, at-home testing of one's hydration level, is cost-effective and thus inline with the sustainable development goals 3 \& 10 of the United Nations, and a step-forward to patient-centric healthcare systems, smart homes, and smart cities of future.

\end{abstract}


%% file: sec1.tex
\section{Introduction}
\label{sec:intro}

Water is a vital component of the human body, making up approximately 60\% of its composition \cite{1d}. This significant proportion emphasizes the importance of water in various physiological functions and in maintaining overall health. The body's daily requirement of roughly 1.5 liters of water is critical for numerous essential roles that water fulfills \cite{2d}. Water's role extends beyond mere hydration; it forms the foundation of new cell membranes \cite{Cooper2000-xq}, a process crucial for cell health and regeneration. Additionally, water enhances the body's metabolic rate, assisting in weight management \cite{Thornton2016-nt}, and is key in transporting proteins and carbohydrates from consumed food \cite{Sahoo2014-fn}, \cite{Lee2015-pl}. It is also instrumental in the elimination of toxins, playing a major role in urination and sweating \cite{Genuis2013-ai}-\cite{Genuis2011-ux}. Water is also vital for brain function \cite{Keep2012-hc}, and for cognitive health.

The significance of water for human body becomes more pronounced when considering hydration states. A person with a total body water (TBW) content of approximately 65\%-70\% is deemed well-hydrated and healthy. Even a minor reduction in TBW can lead to dehydration, impacting bodily functions \cite{EFSA_Panel}. Dehydration is characterized by a negative water balance in the body, where water intake is insufficient to replace the water lost \cite{12d}, leading to an imbalance that can affect many aspects of health and bodily performance altering physiological parameters, and could result in a myriad of health risks.  

The symptoms of dehydration can range from mild, such as headaches, high heart rate, and dizziness \cite{13d}, to severe, including seizures, kidney failure, myocardial infarction, and potentially fatal outcomes \cite{12d}. Notably, even a minor reduction, as little as 1\%-2\% in body water, can have significant effects on both mental and physical performance \cite{16d}. This condition is particularly perilous in the elderly, attributed to their reduced body water content, diminished sense of thirst, and prevalent kidney issues. Studies have shown that dehydration significantly escalates the risk of mortality among hospitalized elderly patients \cite{14d}. Furthermore, certain professions, like soldiers and construction workers who engage in continuous, strenuous labor in extreme weather conditions, are at an elevated risk for dehydration, highlighting the need for efficient hydration monitoring.  Dehydration's impact extends beyond mere physiological distress; it can also contribute to conditions such as obesity in children, stemming from electrolyte imbalances \cite{15d}. 

In the field of dehydration assessment, established methods such as body mass changes and isotope dilution are considered the gold standard. These techniques, while offering accuracy, require sophisticated and costly equipment, as well as controlled procedures. For example, body mass changes require specific conditions to accurately reflect water weight variations apart from muscle or fat mass changes \cite{m1}. Isotope dilution, a method that involves ingesting a known quantity of an isotope and then measuring its concentration in bodily fluids, is accurate but impractical for routine use \cite{m2}. Clinicians also commonly use blood and urine analysis for hydration levels, assessing plasma osmolality or electrolyte balance \cite{m3}, and examining urine's specific gravity or color \cite{m4}, \cite{m5}, \cite{m6}. Additionally, as saliva is always available, salivary osmolality has been proven to be a reliable biomarker for hydration status \cite{m6}. Further, microbiology based methods employ leg skin microbiome data from subjects for precise prediction of skin hydration levels along with various other significant biological markers \cite{carrieri2020explainable}. However, these diagnostic methods necessitate advanced and expensive instrumentation, in-depth laboratory analysis, and specialized medical staff. Consequently, they are more apt for occasional use rather than frequent monitoring, often leading to their employment in emergency scenarios, such as when patients are admitted to hospitals with advanced stages of dehydration.

Keeping in mind the limitations of the traditional methods of dehydration assessment, there's a rising demand for solutions that are portable, non-invasive, and user-friendly. Smartphones, widely available and equipped with sufficient sensing and computational capabilities, have recently received attention a potential tool for in-situ medical diagnostics \cite{ahsannaturepaper}. Smartphones not only offer quick and accessible diagnostics, bypassing the lengthy waiting periods associated with hospital-based tests, but also provide a more cost-effective alternative for regular hydration assessments compared to traditional laboratory methods and equipment.

\subsection{ Contributions } 
We propose to utilize the camera of a regular smartphone to record a small video of the fingertip for non-invasive dehydration monitoring. 

{\it Rationale:}
Dehydration could reduce the overall blood volume and increase blood viscosity, which may in turn increase the heart rate and reduce the blood pressure. This motivates us to infer dehydration related information by collecting video data from fingertip of the subject, which contains the photoplethysmography (PPG) signal that captures the variations in the peripheral blood volume due to change in hydration level of a person over time. 

The key contributions of this work are as follows:   
\begin{itemize}
    \item We have acquired a novel labeled video-PPG dataset for dehydration assessment, first of its kind. That is, we have constructed a custom multi-session labeled dataset by collecting 125 minutes worth of video-PPG data from 25 fasting subjects during the month of Ramadan in 2023. It is worth mentioning that each subject fasted for a duration of 13 hours (from 6 am till 7 pm).
    \item We propose a binary classification framework, i.e., dehydration detection, using the pre-processed and augmented PPG time-series data. A large number of our machine learning (ML), deep learning (DL) and transformer models reliably determine whether a person is hydrated or dehydrated.
    \item We propose a 4-class classification framework, i.e., our ML, DL and transformer models reliably rate the dehydration level of a person on a scale of 1 to 4, again using the pre-processed and augmented PPG time-series data. 
    \item We propose another low-complexity method whereby we feed high-dimensional PPG time-series data to a DL model for feature extraction, followed by t-SNE method for feature selection and dimensionality reduction, followed by a number of ML classifiers that do dehydration level classification in a reliable manner.
    \item Finally, we interpret the decisions by our best-performing DL model by identifying the most relevant features in the PPG time-series data under the SHAP-based explainable AI framework.
\end{itemize}
%Fig. \ref{fig:methodology} shows a pictorial summary of the proposed methodology.

{\it To the best of our knowledge, this is the first work that: 1) utilizes a smartphone camera to do non-invasive dehydration monitoring, 2) does dehydration level classification on a scale of 1 to 4. Further, the proposed method (with a maximum accuracy of 99.65\%) outperforms all the previous works on non-invasive dehydration monitoring in the literature where maximum reported accuracy is 97.83\%. }

The proposed method allows rapid, do-it-yourself, at-home testing of one's hydration level, is cost-effective and thus inline with the sustainable development goals 3 \& 10 of the United Nations, and a step-forward to patient-centric healthcare systems, smart homes, and smart cities of the future.

\begin{figure*}[!t]
\centering
\includegraphics[width=\textwidth]{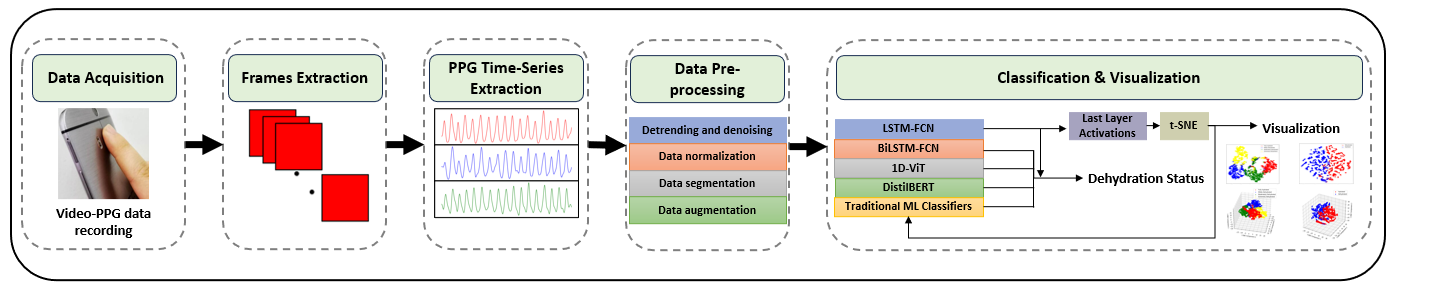} % Replace with your figure file
\caption{Pictorial overview of our proposed smartphone-based method for non-invasive dehydration monitoring.}
\label{fig:methodology}
\end{figure*}

% 1) We propose hand-breathe, a novel non-contact method to detect breathing abnormalities. Specifically, the subject rests his/her hand on a table in between the transmit and receive horn antennas. The transmitted OFDM signal passes through the hand, and is collected at the receive end. The receiver extracts from the received signal the channel frequency response (CFR), i.e., fined-grained wireless channel state information (WCSI) which is then fed to various ML classifiers which eventually classify between different breathing abnormalities. Among all classifiers, linear support vector machine (SVM) classifier yielded the best accuracy which is 88.1\%.  

% Note that the proposed method utilizes the OFDM signal because of its wideband sensing capability. Specifically, each of the $N$ subcarriers of the OFDM provides distinct information about the modulation of the RF signal reflected-off the human body. Thus, the received OFDM signal helps us observe the effect of respiration-induced near-periodic motion of the chest and the hand, over a wide range of frequencies which eventually helps our ML classifiers achieve a superior classification accuracy.  

%One key benefit of the proposed platform is that it is portable, scalable, flexible, and low cost (and thus, quite amenable for rapid testing of masses for covid19).  

\subsection{Outline} 
The rest of this paper is organized as follows. Section II discusses the related work. Section III provides sufficient details about our data collection process. Section IV summarizes the key data pre-processing steps. Section V provides essential details about the various ML, DL, and transformer models we have implemented in Matlab and Python. Section VI provides a detailed performance analysis of our AI models. Section VII presents data visualization, an alternate method for dehydration monitoring, and interpretation of the decisions by our DL model under the SHAP-based explainable AI framework. Section VIII concludes the paper. 

\section{Related Work}

Very recently, there has been an increasing interest to design novel non-invasive methods for dehydration monitoring, in addition to the traditional biochemistry based methods that rely upon blood, saliva, sweat or urine samples. Since this work is positioned as a non-invasive method; therefore, the related work closest to the scope of our paper include the papers whereby authors employ wearable sensors, e.g., pulse oximeters, smartwatches, smart wristbands, smartphones, smart tattoos/stickers, for non-invasive dehydration monitoring. Alternatively, viewing from a different lens of sensing modalities, the related work attempts to infer dehydration level of a person using bio-signals, e.g., PPG, electrodermal activity (EDA), bioimpedance sensors, etc., by passing them through various ML and DL algorithms. Below, we discuss the related work by categorizing it according to the various sensing modalities authors have used in their research.

{\it EDA-based methods:}
In \cite{liaqat2020non}, researchers collected EDA data from five subjects using the BITalino toolkit to assess hydration levels under two scenarios: hydrated and dehydrated states. The data was further categorized based on physical postures such as sitting, standing, and lying down. Participants were classified as dehydrated after fasting for at least ten hours and as hydrated if they had consumed water regularly. Employing feature extraction techniques and the Random Forest algorithm, the study achieved an  accuracy of 91.3\% in predicting the hydration status. In a similar work \cite{liaqat2022personalized}, the authors collected data from 16 participants in three hydration states: hydrated, mildly dehydrated, and extremely dehydrated. The data was gathered in different body postures using the EDA sensor and BITalino toolkit. They focused on extracting key statistical features from the data in order to feed it to their hybrid BiLSTM that ultimately achieved an accuracy of approximately 97\%. Further, in \cite{rizwan2020non}, the authors collected EDA data from several subjects in different body postures, such as sitting and standing, under varying hydration conditions. The study utilized the BITalino EDA sensor for data collection and implemented feature extraction, focusing on statistical features such as mean, variance, and kurtosis. To classify the hydration status, six ML algorithms were evaluated, with the K-Nearest Neighbour (K-NN) algorithm achieving the highest accuracy of 87.78\%. In \cite{kulkarni2021non}, researchers developed an Android app titled {\it monitoring my dehydration} which harnessed EDA data collected via a wristband and fed it a convolutional neural network (CNN) for binary classification of the hydration level of a person. Their app factored in four physical activities--walking, sitting, standing, and lying down--along with two hand gestures. When tested with data from five individuals, their app achieved a classification accuracy of 84.5\%. 
Finally, \cite{S2023SmartphoneBS} proposed a hydration monitoring method by utilizing EDA signals, which were analyzed using the dynamic time warping (DTW) algorithm. Their proposed architecture includes the integration of a microcontroller and a cloud-based database for the storage and analysis of data. Additionally, they developed a mobile application for real-time monitoring of hydration status.

{\it PPG-based methods:}
\cite{reljin2018automatic} developed an automated, non-invasive approach to detect dehydration using PPG signals. The researchers gathered data from key body locations-- the forehead, ear, and finger-- of 42 patients suffering from dehydration. By analyzing specific characteristics of the PPG signals, including the goodness of fit and changes in amplitude within the heart rate frequency range, they employed support vector machines (SVM) with a radial basis function (RBF) kernel for the classification task. Their method ultimately achieved a rather low classification accuracy of 67.91\%.  

{\it PPG \& EDA-based methods:}
In \cite{posada2019mild}, researchers conducted a study to detect mild dehydration by collecting both EDA and PPG data from 17 subjects. EDA was measured using stainless-steel electrodes on the left hand connected to a galvanic skin response module, and PPG signals were recorded with a wearable device on the left wrist. The analysis was performed using various ML algorithms, focusing on autonomic responses to cognitive stress induced by the Stroop test. The authors were able to detect mild dehydration with an overall accuracy of 91.2\% using SVM technique. In the research presented in \cite{suryadevara2015towards}, data including EDA, skin temperature, PPG readings, and body mass index were gathered from 16 participants during circuit training sessions. This information was subsequently analyzed using a specially derived empirical formula to determine the level of fluid loss, a key indicator of dehydration, that occurred due to the exercise. \cite{Sabry2022} proposed a regression-based approach for dehydration monitoring, utilizing data from  EDA and PPG sensors. This method focuses on predicting the last drinking time by analyzing readings from these sensors. Their approach also integrates additional data from wearable sensors like accelerometers, magnetometers, gyroscopes, and sensors for temperature and barometric pressure, to comprehensively assess hydration status.

{\it ECG-based method:}
The study in \cite{alvarez2019machine} utilized 12 lead-ECG signals from the Cardiosoft device in a clinical setting to classify three hydration stages during exercise using the SVM classifier. They took measurements before exercise, post-exercise, and after hydration. They identified the RR-interval as a key feature which helped their SVM classifier achieve a classification accuracy of 60\%. 

{\it Sweat-based \& bioimpedance-based methods:}
In a series of studies, Ring et al. have investigated various non-invasive methods to estimate total body water (TBW) loss, exploring different approaches in each work. In \cite{7320006}, the focus was on the viability of using sweat electrolyte concentrations, specifically chloride and osmolality, as estimators for TBW loss. Utilizing sweat collectors during exercise, the study found significant correlations between TBW loss and sweat chloride concentration (r = 0.41) and osmolality (r = 0.43).
Building on this work, the researchers proposed a method in \cite{7182268} to enhance the accuracy of TBW estimation by adjusting bioimpedance measurements for temperature effects. Here, they employed ML techniques, including linear and nonlinear support vector regression and Gaussian process regression, to effectively mitigate the impact of temperature on bioimpedance readings.
Continuing their exploration in \cite{7539317}, the researchers then turned to salivary markers as a tool for estimating TBW loss. They aimed to utilize various substances found in saliva, such as electrolytes and hormones, applying ML algorithms to refine the accuracy of hydration loss measurement in physical activities. This approach, which utilized techniques like Gaussian process regression and linear regression with k-exhaustive feature selection, achieved an estimation error of about 0.34 liters.

{\it Acoustic-based methods:}
In their study \cite{7759295}, the authors introduced the AutoHydrate system, which employs acoustic signals captured by a throat microphone. This system extracts time and frequency domain features, which are then analyzed using an SVM classifier to determine drinking activities with a classification accuracy of 91.5\%. Additionally, the processed data is transmitted via Bluetooth to a smartphone application, facilitating the display of hydration monitoring results.

{\it Smartphone-based methods:}
In \cite{sm13}, the authors introduce a smartphone-based tool for non-invasive dehydration detection, utilizing a dataset of 2,340 facial images and a siamese neural network model to distinguish the two hydration levels, demonstrating an accuracy of 76.1\%. Also, in \cite{sm14} smartphones were employed to record skin turgor tests, utilizing two techniques--skin marking and texture analysis. Subsequently, the authors utilized image processing algorithms to analyze video frames to derive skin mechanical characteristics and hydration status, simulating the traditional medical approach of observing skin's elasticity and recovery during dehydration assessments. 

{\it Non-contact methods:} 
Diverging from non-invasive methods, the authors in \cite{hasanhydrationpaper} have developed a unique non-contact method for monitoring dehydration through the use of radio frequency (RF) signals. Utilizing software-defined radios (SDRs), their system directs RF signals towards the chest or hand of a subject and then captures the reflected signals to determine hydration status. Their study involved data collection from five fasting subjects during the month of Ramadan. With this data, their neural network classifier yielded an accuracy of 93.8\% for their chest-based method and 96.15\% for their hand-based method.

%% file: sec3-1.tex
\section{Data Acquisition}

% 1) Subjects: Ten male subjects (179 ± 7.5 cm; 79.3
% ± 9.0 kg; 25.5 ± 3.7 years) volunteered to participate in
% the study. All subjects provided written informed consent
% after the study protocol was approved by the local ethics
% committee.

This section describes the first major contribution of this work, i.e., a novel labeled video-PPG dataset for dehydration assessment in a non-invasive manner. 

{\it The need for the new dataset:}
The data acquisition campaign aimed to construct a custom video-PPG dataset, driven by the absence of publicly available labeled video-PPG datasets suitable for dehydration assessment (see Table \ref{table:public_datasets}). This lack of availability of public datasets on non-invasive dehydration monitoring is in turn due to the fact that though the acquisition of the raw video-PPG data is relatively easy and straightforward, yet it is difficult to label this data except in the clinical setting where a trained medical professional obtains blood, urine, or saliva samples of the subject and goes on to do bio-chemistry analysis to obtain the relevant dehydration related biomarkers, e.g., plasma/urine osmolality, urine specific gravity, etc.

{\it Data labeling strategy:}
Since we conduct the data collection on the premises of an academic campus (instead of a hospital), we adopt a rather innovative approach to label the video-PPG data. Specifically, we chose to collect data from a control group of fasting subjects during the month of Ramadan in 2023. In our study, subjects consistently fasted for approximately 13 hours each day (typically from 6 am to 7 pm), during a season characterized by high temperatures, indicative of the hot weather conditions prevalent during that specific time frame. This control group of fasting subjects allowed us to label the video-PPG data collected from the fasting subjects in a systematic manner\footnote{This research study was approved by the local ethical institutional review board of King Abdullah University of Science and technology (KAUST). All subjects provided their written informed consent before the data collection.}. That is, for the binary classification problem of dehydration detection, the data collected an hour before (after) the sunset/breaking of fast was labeled as belonging to dehydrated (hydrated) class. 

{\it Data labelling strategy for multi-class classification problem:}
For the multi-class classification, where the objective is to rate the dehydration level of a person on a scale of 1 to 4, we capitalize on the fact that the hydration level of a fasting person monotonically decreases from morning till evening. This allows us to collect five distinct measurements (at regular intervals) from each fasting subject between sunrise and sunset. Furthermore, we label the raw video-PPG data obtained during the five measurements as follows:
\begin{itemize}
    \item {\it Level 1--Fully Hydrated (FH):} The first measurement is taken at the beginning of the day, when the fasting subjects are maximally hydrated.
    \item{\it Level 2--Mildly Dehydrated (MD1):} The second measurement is taken 2-3 hours later, at which point the hydration levels of the fasting subjects have decreased slightly.
    \item{\it Level 3--Moderately Dehydrated (MD2):} The third measurement is taken another 2-3 hours later, as the hydration levels of the subjects continue to decrease.
    \item{\it Level 4--Extremely Dehydrated (ED):} "The fourth measurement is taken after another 2-3 hours, essentially an hour before the subjects break their fasts, at which point the subjects are maximally dehydrated.
    \item{\it Rehydrated (RH):} The fifth measurement is taken one hour after sunset, after the subjects break their fasts, thus having the opportunity to rehydrate and return to a fully hydrated state.
\end{itemize}

{\it Key statistics of the custom dataset:}
In total, we collect raw video-PPG data from 25 fasting subjects (16 males, 9 females) with an average age of 29.4 years, an average weight of 63 kg, and an average height of 164.23 cm. We use an android Vivo phone with a camera frame rate of 30 Hz. During each measurement session, each subject placed his/her fingertip on the rear camera of the smartphone in order to let us record the video-PPG data for a duration of 5 minutes. This implies that we collect $25\times 5=125$ minutes $=7,500$ seconds worth of video PPG data for each data class/session. This in turn implies that we have $7,500\times 2=15,000$ seconds worth of labeled data for binary classification, and $7,500\times 5=37,500$ seconds worth of labeled data for multi-class classification. Last but not the least, we also collect meta-data from each subject, e.g., age, weight, height, gender, etc.
\begin{table}[t!]
\centering
\caption{A quick summary of publicly available datasets for non-invasive dehydration monitoring}
\label{table:public_datasets}
\setlength{\tabcolsep}{2pt} % Reduce horizontal padding
\renewcommand{\arraystretch}{1} % Adjust vertical spacing
\footnotesize % Use smaller font size
\begin{tabular}{|c|p{2.8cm}|p{2.8cm}|c|} 
\hline
{\bf Work} & {\bf Device} & {\bf Modality} & {\bf Subjects} \\
\hline
\cite{s22051887} & Shimmer3 GSR & accelerometer, gyroscope, temperature, magnetometer, EDA, pressure, PPG & 11 \\
\hline
\cite{Ring2017-ua} &  Bioimpedance device, laboratory equipment, thermometer, sweat collector & bioimpedance, temperature, salivary samples, sweat samples & 10 \\
\hline 
\cite{Kirby2021-jn} & Smartwatch (SpectroPhon) & total salt in sweat, sweat mass & 240 \\
\hline
\cellcolor{green!25} This work & \cellcolor{green!25}Smartphone & \cellcolor{green!25}video-PPG & \cellcolor{green!25}25 \\
\hline
\end{tabular}
\vspace{-1mm}
\end{table}

% \begin{figure}[ht]
% \begin{center}
% 	\includegraphics[width=5cm,height=5cm]{Figures/watch.png} 
% \caption{The proposed non-contact method for dehydration monitoring: The apparatus consists of an SDR-based RF transceiver to collect radio data off the chest and the hand of the subject. The collected data is subsequently passed to various machine learning methods, which ultimately classify a subject either as hydrated or dehydrated.}
% \label{fig:sysmodel}
% \end{center}
% \end{figure}

% \begin{figure}[!ht]
%     \centering
%     \subfloat[Empatica Embrace Plus ]{\includegraphics[width=0.45\linewidth]{Figures/watch.png}}\hfill
%     \subfloat[VIVO smartphone]{\includegraphics[width=0.45\linewidth]{Figures/watch.png}}
%     \caption{The tools used in collcted the dehyration reletaed data. (a) EDA data collection using Empatica Embrace Plus  done by placing the watch on the wrist. (b) PPG signal was obtained by placing the index finger on the VIVO smartphone's rear camera }
%     \label{fig:combined}
% \end{figure}

%% file: sec4-1.tex
% \section{Data Pre-processing \& Training of Neural networks} 

\section{Data Pre-processing} 

Having acquired the custom dataset, we perform a number of pre-processing steps in order to condition the raw video-PPG data and in order to prepare it for the ML, DL and transformer models. Fig. \ref{fig:methodology} shows a pictorial summary of the proposed methodology. Below, we describe the key data pre-processing steps, one by one:
\begin{itemize}
    %\item The video signal duration for each subject was 300 seconds. With a frame rate of 30 frames per second, the 300-second video accumulated a total of 9000 frames. These frames included Red-Green-Blue (RGB) components, resulting in each frame being composed of three channels.
    \item {\it Extraction of PPG time-series:} We begin by doing pixel-averaging for each frame, for each of the three color channels (i.e., red, green and blue) in each video-PPG snippet. This approach allows us to obtain three PPG signals at a sampling rate of 30 samples/sec corresponding to the three color channels, for each measurement session (data class).
    
    \item {\it Removal of artifacts, detrending and denoising:} The three PPG signals (corresponding to three color channels) contained occasional artifacts (due to motion) and baseline (due to respiration), which were removed using discrete wavelet transform-based decomposition and reconstruction method \cite{rabiasensorsletterspaper}. The detrended PPG signals were then passed through a fourth-order Butterworth low-pass filter with cut-off frequency 10 Hz, to remove any residual high-frequency and out-of-band noise. 
    \item {\it Data normalization:} The conditioned PPG time-series data was normalized using z-score normalization method.
    \item {\it Data segmentation:} Subsequently, we segment the conditioned PPG data--with the aim to increase the number of examples in the dataset. Specifically, we apply the windowing method (with a rectangular window of size 3 seconds) to partition each of the conditioned PPG time-series/examples into smaller, non-overlapping segments/examples of duration 3 seconds each. Then, keeping in mind that dehydration is a slowly-varying phenomenon which doesn't change over a small duration of a few cardiac cycles, we replicate the class label of each original unsegmented PPG time-series to each of its segments. This way, the data segmentation step increases the number of examples per class to 2,500, which later helps the AI models learn better the dehydration-related features in the dataset.
    \item {\it Data augmentation:} To further enhance the ability of the AI models to generalize, we apply Gaussian noise-based data augmentation. Specifically, we utilize zero-mean Gaussian noise with a small standard deviation of 0.01 to contaminate the dataset. This approach is used to increase the size of the segmented dataset by a factor of two, three, and four, respectively.\footnote{{Additive Gaussian noise-based data augmentation is a well-known method to introduce realistic perturbations in the data. The added noise has zero mean so that the augmented data could retain the characteristics of the original data, while a standard deviation of 0.01 introduces realistic variability in the augmented data. Thus, data augmentation helps improve the generalization capabilities of the developed AI models, without distorting the intrinsic signal patterns.}}. The data augmentation not only increased the size of the dataset, but also introduced necessary variability that mimics real-world PPG signal fluctuations, thereby enhancing the robustness of the AI models. 

\end{itemize}

% \begin{figure}[ht]
% \begin{center}
% 	\includegraphics[width=9cm,height=4cm]{Figures/PPG Digram.png} 
% \caption{PPG signal Extraction.}
% \label{fig:sysmodel}
% \end{center}
% \end{figure}

%This process involved generating an array of Gaussian random variables, matching the size of each data segment. Each segment was then duplicated, with one version remaining original and the other having the Gaussian noise array added to it. This approach effectively doubled our dataset, creating two distinct versions of each segment.

%To further augment the dataset by a factor of three, a second set of Gaussian random variables was generated and added to the original segments, resulting in three unique versions: the original, the first noise-augmented, and the second noise-augmented. A similar process was repeated to achieve a fourfold increase in the dataset size, each segment having four versions - the original and three distinct noise-augmented variations. 

%% file: sec4-2.tex
\section{Machine Learning, Deep Learning and Transformer Models Implemented}

We train and evaluate a wide range of machine learning, deep learning, and transformer-based methods for dehydration level classification. This includes 1) binary classification (hydrated vs. dehydrated), and 2) 4-class classification (fully hydrated, mildly dehydrated, moderately dehydrated, and extremely dehydrated).

\subsection{Machine learning models}

We utilize the Matlab's Classification Learner app to implement a plurality of ML methods, e.g., K-nearest neighbours (KNN) with different distance metrics, support vector machines (SVM) with various kernels, ensemble classifiers, kernel logistic regression (KLR), and neural networks (NN). Each method is chosen for its ability to capture different kinds of patterns within the data. More pertinent details about the ML classifiers implemented are as follows:

\begin{itemize}
    \item {\it KNN classifiers:} We utilize the following three different variants of the KNN classifier: KNN-Euclidean (with $k=1$), KNN-Cubic (with $k=1$), and KNN-Cosine (with $k=10$). Note that each of these three KNN variants utilizes a different distance metric. For example, KNN-Cosine classifier considers also the angle (in addition to the distance) between data points in feature space.
    \item {\it SVM and KLR classifiers:} We train a Kernel-SVM classifier with two different kernels, with the aim to project the data into a higher-dimensional space for linear separation. We also utilize the SVM-Fine Gaussian classifier that uses a fine-tuned Gaussian kernel to capture more subtle variations in the data. Further, we also implement Kernel logistic regression method that leverages the power of kernel methods in the probabilistic framework of logistic regression.
    \item {\it Ensemble methods:} We utilize two ensemble methods: Ensemble-Subspace KNN and Ensemble-Bagged Trees, to take advantage of the collective decision-making of multiple models, thereby improving the generalization capabilities of the developed classification system.
    \item {\it Shallow neural network:} We explore three different variants of a standard shallow neural network: 1) '2x10 NN', consisting of two hidden layers with ten neurons each, 2) '1x100 NN', consisting of a single hidden layer with one hundred neurons, and 3) '3x10 NN', consisting of three hidden layers with ten neurons each. This approach allows us to evaluate a shallow neural network for its ability to learn non-linear relationships across different levels of complexity.
\end{itemize}

%Overall, these classifiers were meticulously selected and trained to provide a comprehensive understanding of their efficacy in predicting hydration status in both binary and multi-class scenarios, thereby contributing to the reliability and robustness of our findings.

\subsection{Deep learning models} 

Since a PPG signal represents the temporal dynamics of various physiological processes of the body, it is natural to think of using recurrent neural networks (RNNs) which could effectively process such sequential data (by utilizing their internal states to retain contextual information). 

\subsubsection{Long Short-Term Memory-Fully Convolutional Network (LSTM-FCN)}

%Karim et al. introduced an innovative approach that integrated a long short-term memory recurrent neural network (LSTM-RNN) module into fully convolutional networks (FCNs) for time series classification \cite{Karim2018}. The model exhibited improved classification accuracy without intricate data pre-processing. 

\begin{figure}[ht]
\begin{center}	\includegraphics[width=5cm,height=8cm]{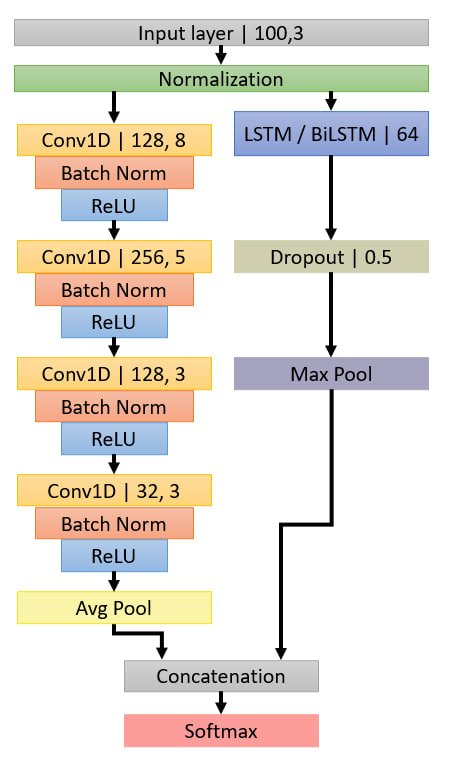} 
\caption{Architectures of the purpose-built deep learning models (LSTM-FCN and BiLSTM-FCN) for dehydration level classification.}
\label{fig:diagram_lstm_bilstm}
\end{center}
\end{figure}

One notably successful variant of RNNs is long short-term memory (LSTM). Instead of a simple LSTM model, we choose to implement an LSTM-FCN model for our problem, as it combines the strengths of both LSTM and FCN architectures for time-series analysis tasks. That is, the LSTM-FCN model is designed to capture both the temporal dependencies in sequential data through the LSTM layers and the local patterns through the convolutional layers. Thus, we train and fine-tune a custom LSTM-FCN model with four convolutional layers and one LSTM layer, followed by a dropout layer, to perform binary and multi-class classification for non-invasive dehydration monitoring using PPG time-series data. Fig. \ref{fig:diagram_lstm_bilstm} presents the architecture of the purpose-built LSTM-FCN model in detail. 

% \begin{figure}[ht]
% \begin{center}	\includegraphics[width=7cm,height=6cm]{Figures/comp_Lstm.png} 
% \caption{LSTM and BiLSTM Architectures.}
% \label{fig:sysmodel}
% \end{center}
% \end{figure}
% \begin{figure}[ht]
% \begin{center}	\includegraphics[width=8cm,height=6cm]{Confusion_Matrix_lstmfcn_binary_crossentropy.png} 
% \caption{Confusion Matrix for Binary classification using LSTM-FCN model with a test accuracy of 95.23\%}
% \label{fig:sysmodel}
% \end{center}
% \end{figure}
% \begin{figure}[ht]
% \begin{center}	\includegraphics[width=8cm,height=6cm]{Figures/Confusion Matrix_bilstm_4class.png} 
% \caption{Confusion Matrix for 4-class classification using LSTM-FCN model with a test accuracy of 87.56\%}
% \label{fig:sysmodel}
% \end{center}
% \end{figure}

\subsubsection{Bidirectional Long Short-Term Memory-Fully Convolutional Network (BiLSTM-FCN)}

We also create a variant of the LSTM-FCN model, termed the BiLSTM-FCN model, by replacing the LSTM layers in the original model with BiLSTM layers. Note that the BiLSTM model itself is a variant of the LSTM model whereby each BiLSTM layer consists of two LSTM layers, one that analyzes the input sequence forward in time and another that analyzes it backward. 
%What sets BiLSTMs apart is their ability to gather information from both previous and subsequent time steps. LSTMs are proficient in handling sequential data by taking into account the past context; however, they process the sequence in only one direction-from the beginning to the end. This uni-directional flow constrains the network to past and present information, leaving future context out of the equation. In contrast, BiLSTMs incorporate an additional layer that processes the sequence in reverse. 
This dual-pathway architecture allows BiLSTMs to capture context from both past and future states at any point in the sequence, which is particularly beneficial  in time-series data such as the PPG data that we consider in this work. Fig. \ref{fig:diagram_lstm_bilstm} presents the architecture of the purpose-built BiLSTM-FCN model in detail. 

% \\BiLSTM model trained over 400 epochs using the cross-entropy loss function, shares similarities with the first model. It performed well in binary classification, distinguishing between fully hydrated and fully dehydrated states in a dataset of 3600 PPG signals, each with 100 data points. It achieved a testing accuracy of 94.22 \% and the corresponding confusion matrix can be found in Figure 7.
% \begin{figure}[ht]
% \begin{center}	\includegraphics[width=8cm,height=6cm]{Figures/Confusion_Matrix_biLSTM_2class.png} 
% \caption{Confusion Matrix for Binary classification using BiLSTM model with a test accuracy of 94.22\%}
% \label{fig:sysmodel}
% \end{center}
% \end{figure}
% For multi-class classification, this model was adapted to classify four hydration states  on a dataset of 9000 examples and it achieved an 86.81\% accuracy. Figure 8 illustrates its performance in this more complex task. 
% \begin{figure}[ht]
% \begin{center}	\includegraphics[width=8cm,height=6cm]{Figures/Confusion Matrix_bilstm_4class.png} 
% \caption{Confusion Matrix for 4-Class classification using BiLSTM model with a test accuracy of 86.81\%}
% \label{fig:sysmodel}
% \end{center}
% \end{figure}

\subsection{Transformer models}

Transformer is a well-known deep learning model in natural language processing (NLP) domain that relies on attention mechanism in order to do sequence-to-sequence conversion, e.g., language translation, text summarization \cite{NIPS2017_3f5ee243}. As a typical sequence-to-sequence model, transformer is built on an encoder-decoder architecture. Unlike its predecessors, i.e., RNNs and LSTMs, which do sequential computation on the data, transformer is uniquely designed for more effective parallel processing which accelerates the training process of transformer. More recently, transformer models have successively been used for analysis of time-series biomedical signals \cite{ahsannaturepaper,Song_Rajan_Thiagarajan_Spanias_2018,8983326,ZEYNALI2023105130,tahir2024cuffless}. Motivated by this, we implement two distinct transformer models in this work for dehydration level classification using PPG time-series data: an NLP transformer known as DistilBERT, and a vision transformer (ViT).

\subsubsection{DistilBERT}

\begin{figure}[ht]
\begin{center}	\includegraphics[width=7cm,height=5cm]{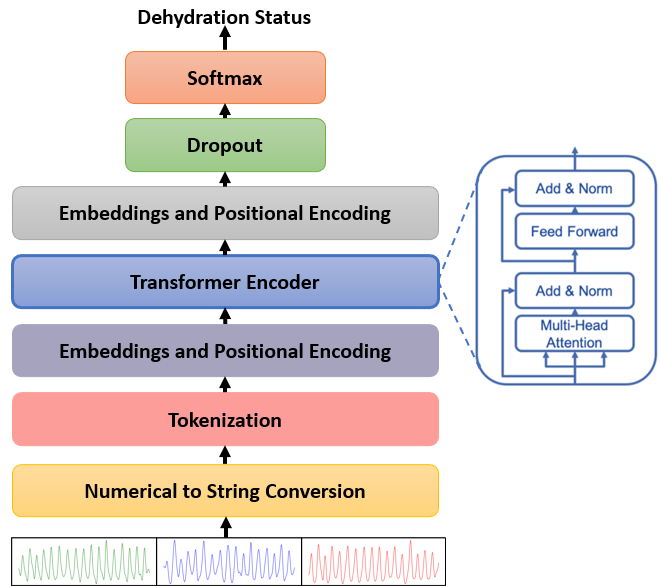} 
\caption{Architecture of the purpose-built DistilBERT model for the downstream task of dehydration level classification.}
\label{fig:distilbertarchitecture}
\end{center}
\end{figure}

BERT (Bidirectional Encoder Representations from Transformers) is an advanced NLP transformer that is pre-trained on a large corpus of text. DistilBERT is a more compact variant of BERT that offers similar performance as BERT but with much fewer parameters which makes it computationally light. Fig. \ref{fig:distilbertarchitecture} illustrates the architecture of the pre-trained DistilBERT model that we implement and fine-tuned for the downstream task of dehydration level classification using PPG time-series data analysis. We begin  by reshaping the three-channel PPG data to a single-channel data (from 100x3 format, to 300x1 format) followed by conversion of data type from numerical to string format, making it suitable for DistilBERT that was originally pre-trained on textual data. Further, in order to do binary and 4-class classification, we utilize only the encoder component of DistilBERT and appended a softmax classification layer at the end, thereby tailoring the model for the classification task while getting rid of the complexity of a decoder. This streamlined approach leverages DistilBERT's efficient feature extraction capabilities, minimizes its computational demands, while making sure that the model determines the dehydration status from PPG signals with high accuracy.

\subsubsection{Vision Transformer (ViT)}

\begin{figure}[ht]
\begin{center}	\includegraphics[width=7cm,height=5cm]{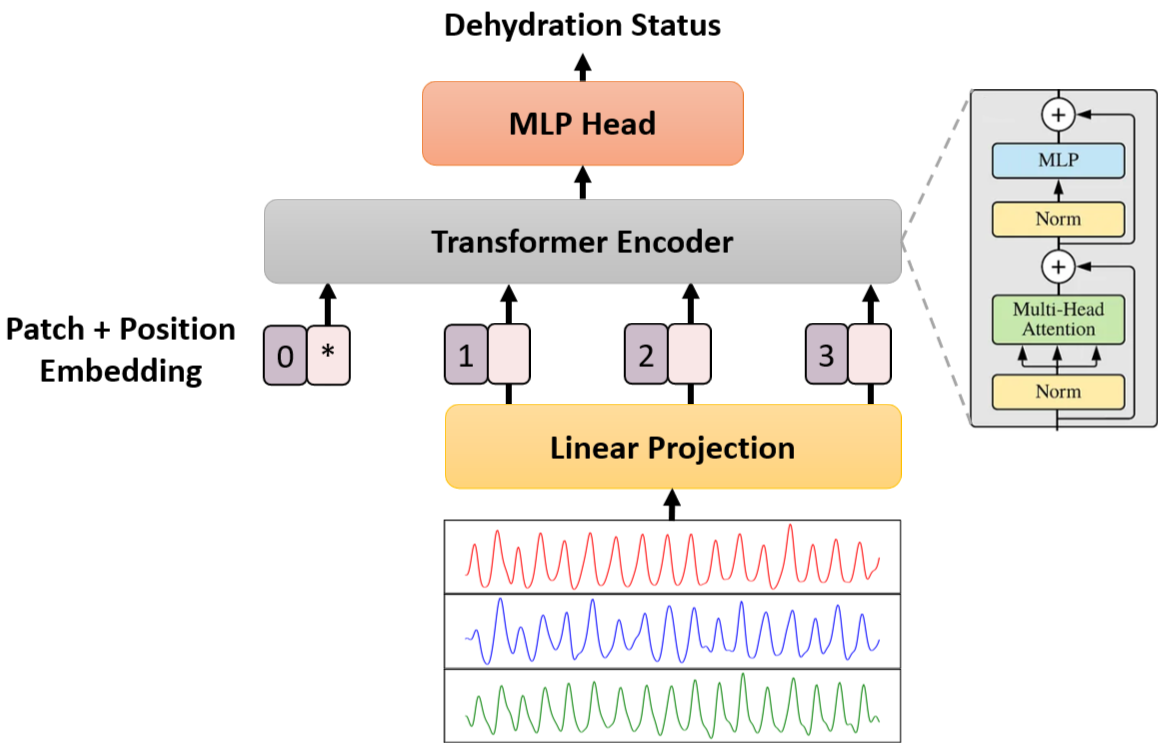} 
\caption{Architecture of the purpose-built 1D-ViT model for the downstream task of dehydration level classification.}
\label{fig:vitmodel}
\end{center}
\end{figure}

The block diagram in Fig. \ref{fig:vitmodel} outlines the architecture of a pre-trained 1D vision transformer (ViT) model that we tailor (by customizing the first few layers of the ViT) for the downstream task of dehydration level classification using PPG time-series data analysis. The PPG signal undergoes a linear projection into flattened patches after initial processing. Positional embeddings are then added to these patches to maintain the sequential information, which is pivotal due to the time-series nature of the PPG data. These embedded patches serve as input to the ViT encoder, which consists of multiple layers, each featuring a deformable attention module and a multi-layer perceptron (MLP) block, with normalization layers interspersed between them. Eventually, the MLP head functions as the classification layer, taking the encoded representations from the ViT encoder and outputting the probabilities of the PPG signal corresponding to different states of dehydration.

\section{Results} 

% \textcolor{red}{it seems probably we are witnessing the overfitting scenario in the app (where the app has wrongly declared me as extremely dehydrated all the times). so this reminds me that we need to include a couple of curves that plot the training and validation loss against epochs. this will help us make the claim in the paper that overfitting is not happening.}

We first list the performance metrics used in this work to evaluate and explain the results, followed by a brief discussion on the hyper-parameters for the developed deep learning and transformer models. We then discuss the performance results of the proposed ML, DL, and transformer models in detail.

\subsection{Performance metrics}
We use three kinds of performance metrics for performance evaluation of the proposed ML, DL and transformer models:
\begin{itemize}
    \item We use confusion matrices, true positive rates (TPR), false positive rates (FPR), true negative rates (TNR), false negative rates (FNR), and accuracy as performance metrics. 
    \item Additionally, we also discuss the flip side of each model, i.e., number of trainable parameters, training time (in seconds), memory requirements (in kB), prediction speed (in observations/sec), number of floating point operations (FLOPs) needed by each model. 
    \item Finally, we use t-SNE (t-distributed stochastic neighbor embedding) method as a tool for dimensionality reduction, and for data visualization. 
\end{itemize}

\subsection{Hyper-parameters of the ML, DL and transformer models}

{\it ML models:} We use a train-validation-test split of 75-15-10 for training and performance evaluation of the proposed ML models and shallow neural network, using Matlab's classification learner app.

{\it LSTM-FCN and BiLSTM models:}
%For binary classification, the model processed a dataset composed of 4500 time-series PPG signals. Each individual signal comprises 100 data points stemming from the RH and ED classes. As for the multi-class classification, The model was retrained to differentiate among the initial four statuses: FD, MD1, MD2, and ED. This subsequent phase introduced a dataset containing a total of 9000 examples, all prepared for input into the LSTM-FCN architecture. 
We use a train-validation-test split of 75-15-10 for training and performance evaluation of the LSTM-FCN and BiLSTM-FCN models in Python+Keras+Tensorflow framework. 

{\it DistilBERT and ViT models:}
We use a train-validation-test split of 75-15-10 for training and performance evaluation of the two transformer models in Python+Keras+Tensorflow framework. An important hyper-parameter of the 1D-ViT model is the patch size. In our implementation of 1D-ViT, we use a patch size of 33 samples. This selection of patch size is based on the fact that the heart rate lies in the range of 60 to 100 beats per minute, and that the PPG time-series in the dataset is sampled at a sampling rate of 30 Hz. This in turn implies that a cardiac cycle consists of 18 to 30 samples, in the dataset. Thus, setting patch size=33 samples ensures each patch covers at least one full cardiac cycle, which helps us retain the complete morphological details of the PPG signal, which in turn help for the efficient dehydration level classification. Thus, in the implementation, we split a PPG segment into three patches.

Table \ref{table:HyperP} summarizes the other important hyper-parameters for the two deep learning models and the two transformer models, e.g., batch size, loss function and more.

%This approach results in three consistent patches, each encompassing a complete cardiac cycle, with the exclusion of one data point.} 

%By carefully selecting this patch size, we ensure that the morphological features of the PPG waveform are preserved, which in turn help for the efficient dehydration level classification. 

\begin{table}[h!]
\centering
\caption{Hyper-Parameters for the LSTM-FCN and BiLSTM-FCN.}
\label{table:HyperP}
\resizebox{\columnwidth}{!}{
\begin{tabular}{|c|c|c|c|c|} 
\hline
Hyper-Parameter & LSTM-FCN & BiLSTM-FCN & DistilBERT & ViT\\
\hline
Epoch  & 200 & 250 & 12 & 100 \\
\hline
Batch size & 32 & 64 & 32 & 32 \\
\hline
Test Ratio & \multicolumn{4}{c|}{10$\%$} \\
\hline 
Validation Ratio & \multicolumn{4}{c|}{15$\%$} \\
\hline 
{Optimizer}  & \multicolumn{4}{c|}{Adam} \\
\hline 
Loss Function & \multicolumn{4}{c|}{Cross-entropy} \\
\hline 
Learning rate & 1e-3 - 1e-4 & 1e-3 - 1e-4 & 1e-5 & 1e-5 \\
\hline 
\end{tabular}
}
\end{table}

% \begin{figure}[ht]
% \begin{center}	\includegraphics[width=9cm,height=5cm]{Figures/barplot_binary_dL (4).png} 
% \caption{Accuracy comparison of LSTM-FCN, BiLSTM, DistilBERT, and ViT models in a binary classification task under varying data augmentation scenarios. Each bar represents model accuracy under (1x) no data augmentation, 2x, 3x, and 4x augmentation factors}
% \label{fig:sysmodel}
% \end{center}
% \end{figure}

% \begin{figure}[ht]
% \hfill 
% \begin{center}	\includegraphics[width=7cm,height=6cm]{Figures/c_m1.png} 
% \caption{confusion matrix for binary dehydration classification, comparing true and false predictions across models.}
% \label{fig:sysmodel}
% \end{center}
% \end{figure}

% \begin{figure}[ht]
% \hfill 
% \begin{center}	\includegraphics[width=7cm,height=6cm]
% {Figures/c_m_2.png} 
% \caption{confusion matrix for classifiers in 4-class dehydration classification, detailing accuracy percentages for each category.}
% \label{fig:sysmodel}
% \end{center}
% \end{figure}

%%%%%%%%%%%%%%%%%
% \begin{figure*}[!t]
% \centering
% \subfigure[First subfigure]{%
%   \includegraphics[width=.45\linewidth]{Figures/mc_ml1.png}%
%   \label{fig:f1}
  
% }
% \hfill
% \subfigure[Second subfigure]{%

%   \includegraphics[width=.45\linewidth]{Figures/4.png}%
%   \label{fig:f2}
% }
% \caption{Two figures side by side across both columns}
% \label{fig:side_by_side}
% \end{figure*}

\subsection{Performance Evaluation of Machine Learning Models}

We analyze the performance of the various machine learning classifiers for hydration level classification using PPG time-series data acquired via a smartphone camera. 

\begin{figure*}[htb]
\centering
% \vspace{-4mm}
\subfigure[Accuracy comparison of the various ML classifiers for binary classification (hydrated vs. dehydrated).]{%
  \includegraphics[width=0.85\linewidth]{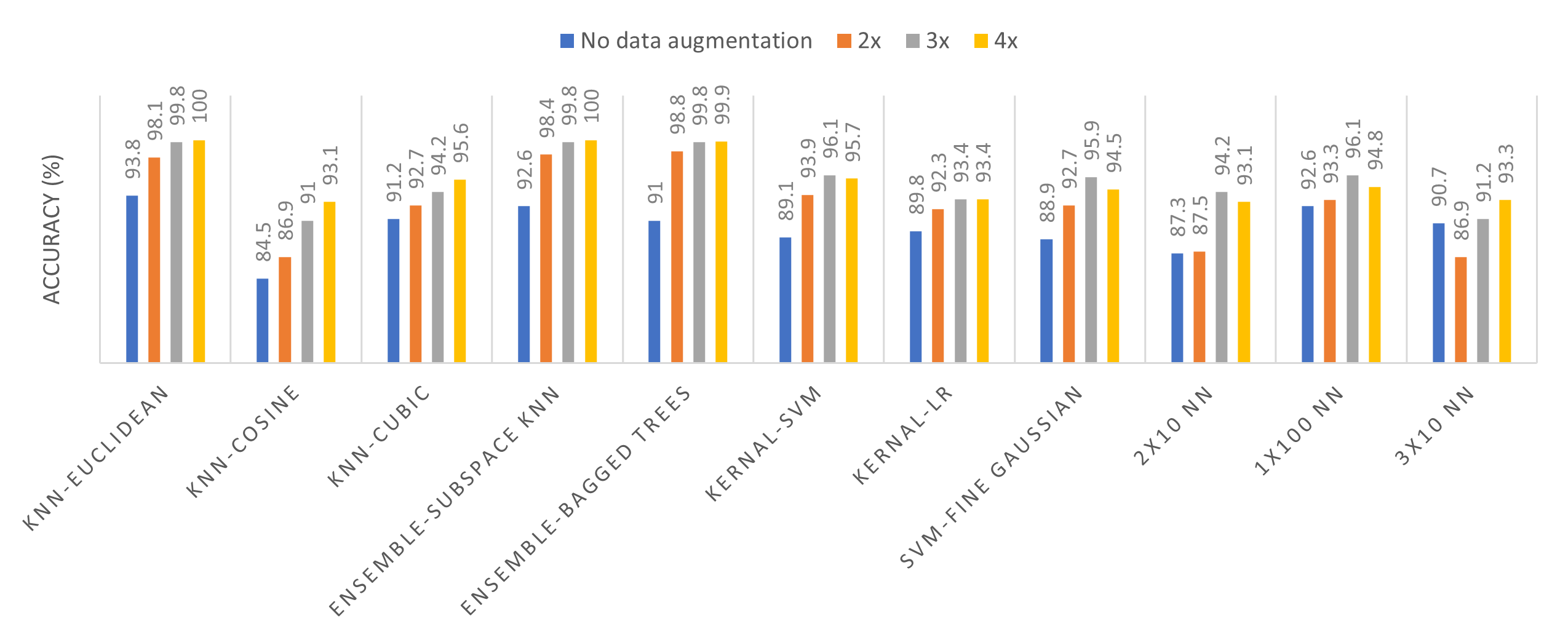}%
}
\hfill
% \vspace{-2mm} % Adjust this value as needed
\subfigure[Accuracy comparison of the various ML classifiers for 4-class classification (fully hydrated, mildly dehydrated, moderately dehydrated, extremely dehydrated). ]{%
  \includegraphics[width=0.85\linewidth]{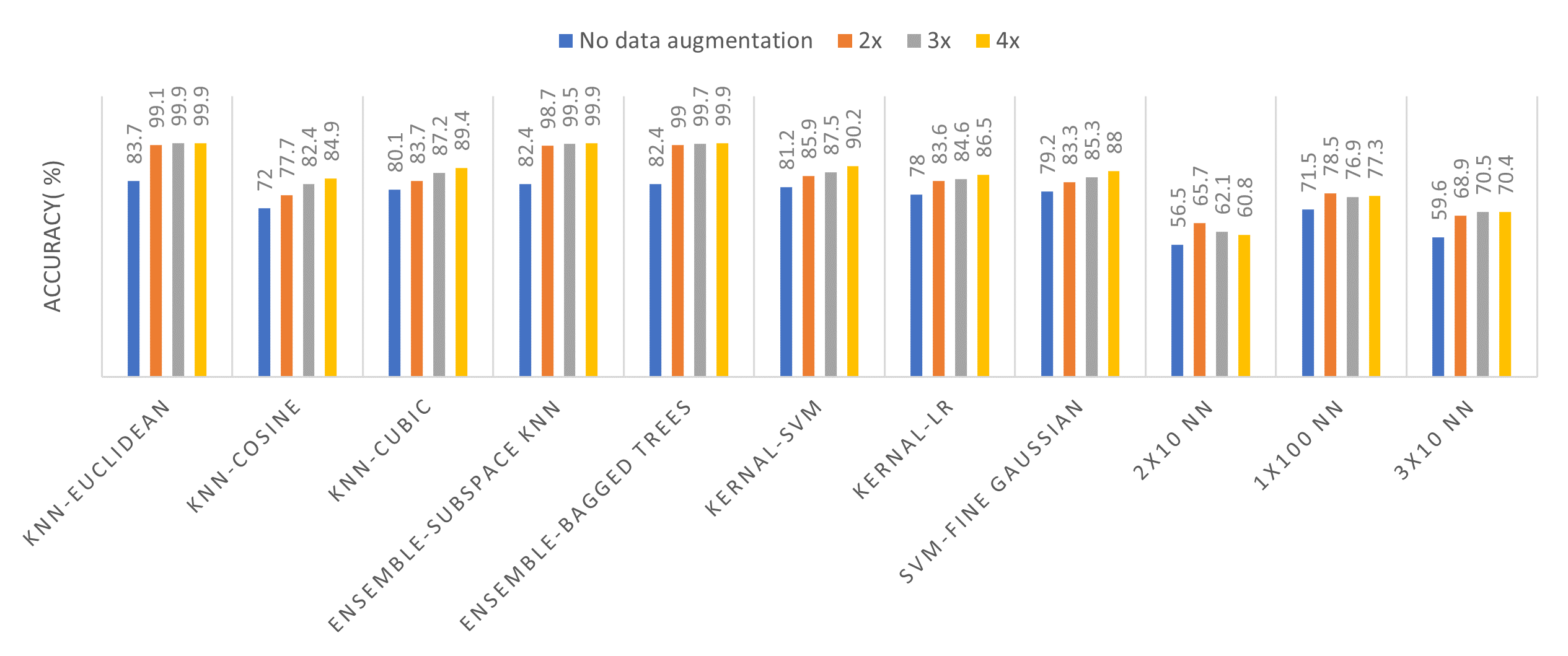}%
}
\caption{Accuracy comparison of the various ML models, i.e., KNN-Euclidean, KNN-Cosine, KNN-Cubic, Ensemble-Subspace KNN, Ensemble-Bagged Trees, Kernel-SVM, Kernel-LR, SVM-Fine Gaussian, 2x10 NN, 1x100 NN, and 3x10 NN classifiers for hydration level classification. Sub-fig. (a) compares the accuracy of the ML models for binary classification, while sub-fig. (b) compares the accuracy of the ML models for 4-class classification. For both sub-figures, we also evaluate the impact of data augmentation (by a factor of 2, 3, and 4) on the accuracy of the ML models.}
\label{fig:accuracyML}
\end{figure*}

We begin with the results for the binary classification problem of dehydration detection (i.e., hydrated vs. dehydrated). Fig. \ref{fig:accuracyML} (a) provides a comprehensive accuracy performance comparison of the various ML classifiers that we implement in this work. We make the following observations from Fig. \ref{fig:accuracyML} (a): 
\begin{itemize}
    \item When there is no data augmentation, the accuracy reported by all the classifiers is still high and remains in a modest range of 84.5\% to 93.8\%, with KNN-Euclidean (KNN-Cosine) classifier performing the best (worst). This points to the pleasant fact that video-PPG (i.e., smartphone)-based dehydration detection is indeed a feasible problem (and not an ill-posed problem). 
    \item An increase in the data augmentation factor (from 2 to 4) leads to (mostly) an increase in the accuracy of all the ML classifiers. However, there is a caveat which we explain next. Consider KNN-Euclidean, Ensemble-subspace KNN, and Ensemble bagged tree classifiers which report a maximum accuracy of 100\% (which is probably overfitting due to data memorization), for a data augmentation factor of 4. Thus, we conclude that doing excessive data augmentation beyond a certain factor results in overfitting. Furthermore, as can be seen in Fig. \ref{fig:accuracyML} (a), the optimal data augmentation factor is different for different classifiers. For example, the optimal data augmentation factor for KNN-Euclidean, Ensemble-subspace KNN, Ensemble bagged tree classifiers is 2, while for some other classifiers, e.g., KNN-Cosine, KNN-Cubic, it is 4. The takeaway message is that the data augmentation (if not excessive) helps us design robust ML classifiers that could efficiently differentiate between meaningful patterns (related to dehydration) and the noise. 
    %\item The more complex 3x10 NN architecture did not consistently outperform the simpler 1x100 NN structure, suggesting that, for our specific dataset and classification challenge, additional complexity in neural network layers may not translate into better performance. A well-calibrated, simpler model might be more effective or even preferable. 
\end{itemize}

%classifiers such as KNN (except KNN-cubic) and Ensemble models exhibit marked accuracy improvements with data augmentation, surpassing enhancements seen in  NN, SVM , and KLR models. For example, the accuracy of the KNN-Euclidean model improved from 93.8\% without augmentation to 98.1\% with data doubled, and further to 100\% when data was quadrupled. The Ensemble-Bagged Trees model saw a similar rise, from 91\% to 99.9\% with the highest level of augmentation.In contrast, the KLR model demonstrated a moderate increase, from 88.9\% without augmentation to 94.9\% with a fourfold increase in data. While NN models also experienced enhancements in accuracy, the improvements were less pronounced; the 1x100 NN model, for example, reaching a peak accuracy of 96.1\% with a tripling of data, up from an initial 92.6\%. 

Next, Fig. \ref{fig:accuracyML} (b) provides a detailed accuracy performance comparison of the various ML classifiers, for the 4-class classification problem (i.e., fully hydrated, mildly dehydrated, moderately dehydrated, fully dehydrated). We learn the following from Fig. \ref{fig:accuracyML} (b): 
\begin{itemize}
    \item First of all, without any data augmentation, the accuracy reported by all the ML classifiers (except the shallow neural network) is modest, and lies in the range of 72\% to 83.7\%, with KNN-Euclidean (KNN-Cosine) classifier performing the best (worst). This again corroborates our hypothesis that video-PPG (i.e., smartphone)-based dehydration classification on a scale of 1 to 4 is indeed feasible (as the four classes are separable, and not entangled).
    \item We again observe that the optimal data augmentation factor is different for different classifiers. For example, the KNN-Euclidean, Ensemble-Subspace KNN, and Ensemble-Bagged Trees models again exhibit potential overfitting as their accuracy reach 100\% for a data augmentation factor of 4. On the contrary, KNN-Cosine, KNN-Cubic, Kernal-SVM, Kernal-LR, SVM-Fine Gaussian show a monotonic increase in accuracy with increase in data augmentation factor, and achieve maximum accuracy when data augmentation factor is 4.  
    \item It's noteworthy that all three variants of the shallow neural network models recorded very low accuracies in this task, with the best accuracy being 78.5\% that was achieved by the 1x100 NN model, for a data augmentation factor of 2. This indicates the potential need for further optimization of neural network architectures to fully exploit the benefits of data augmentation in complex classification scenarios.
\end{itemize}

\begin{figure*}[htb]
\centering
% \vspace{-4mm}
\subfigure[Binary classification]{%
  \includegraphics[width=.45\linewidth]{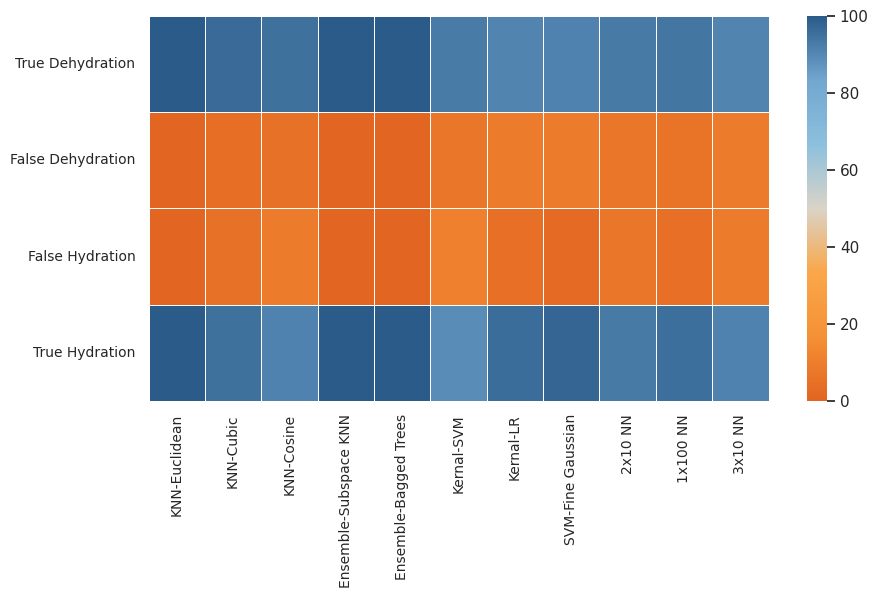}%
}
\hfill
\vspace{-2mm} % Adjust this value as needed
\subfigure[4-class classification]{%
  \includegraphics[width=.45\linewidth]{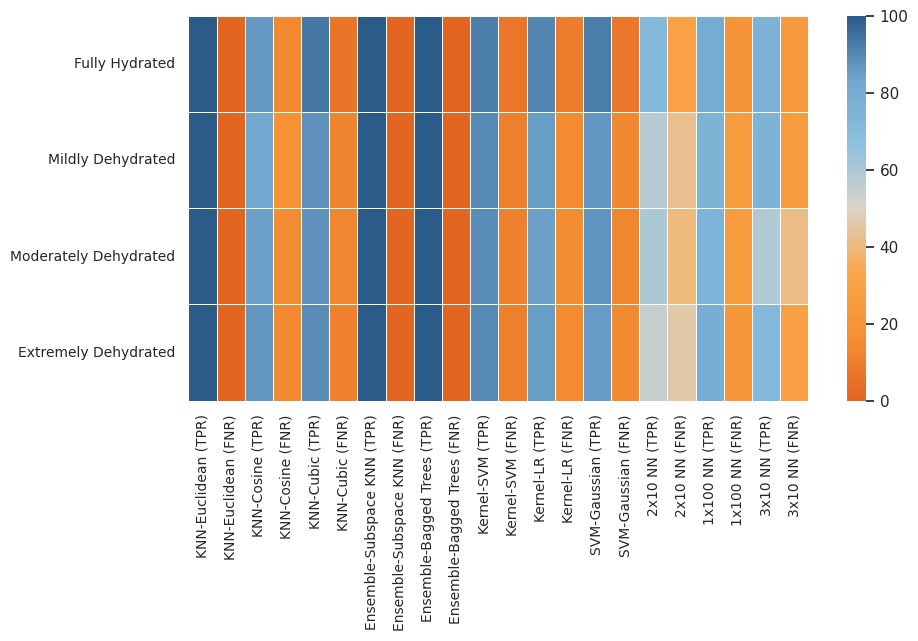}%
}
% \vspace{7mm} % Compensate for the space reduced above
\caption{Heatmaps (confusion matrices) of various ML Classifiers for hydration level classification. Sub-fig. (a) shows the confusion matrices of various ML models for binary classification problem, showing explicitly their true and false predictions (about dehydration and hydration). Sub-fig. (b) shows the confusion matrices of all the ML models for the four-class classification, showing explicitly the four levels of hydration, from fully hydrated to extremely dehydrated. Each heatmap provides a visual representation of the true positive rate (TPR) and false positive rate (FPR) for a given classifier.}
\label{fig:heatmaps}
\end{figure*}

Next, Fig. \ref{fig:heatmaps} (a) (Fig. \ref{fig:heatmaps} (b)) utilizes heatmaps as compact representations of confusion matrices of various ML models that we have implement to do binary (4-class) hydration level classification using the PPG time-series data. Note that this result has been generated using a data augmentation factor of 4. Fig. \ref{fig:heatmaps} is basically an alternate visual illustration of the quantitative results presented in Fig. \ref{fig:accuracyML}, thus corroborating what we infer from Fig. \ref{fig:accuracyML}. For example, in Fig. \ref{fig:accuracyML} (b), the three variants of the shallow neural network are visibly distinguished by their lighter hues, indicating a higher rate of misclassification by the shallow neural networks for both binary and 4-class classification. On the other hand, the darker shades observed for the ML models such as KNN-Euclidean, Ensemble-Bagged Trees, Ensemble- Subspace KNN imply a higher level of accuracy achieved by these models.

%The darker colors in these models' heatmap cells correspond to a greater number of true positives and true negatives, indicating superior performance.

Moving forward, Table \ref{tab:overhead_ML_models} provides a performance comparison of the developed ML models in terms of other alternate performance metrics, e.g., training time, prediction speed, memory needs, and computational complexity (in big-O notation). We learn from Table \ref{tab:overhead_ML_models} that KNN-Euclidean, KNN-Cosine, and SVM-Fine Gaussian models have the smallest training time, while the remaining ML models and the shallow neural network have relatively large training time. Also, we note that all the ML models (except the KNN-cubic and Ensemble subpace-KNN models) are computationally-efficient, as they could make several thousands of predictions every second. Finally, we note that all the ML models (except the Ensemble subspace-KNN model) are light-weight as they require a memory of less than 10 kB. This makes the ML models well-suited for edge computing, i.e., the proposed ML models could easily be implemented as an app that runs on android phones and iPhones (or, even on a low-end modules, e.g., Raspberry PI, Arduino UNO etc.). {Note that we train and evaluat the ML models using Matlab Online that utilizes the Matlab cloud server. Thus, the machine-specific parameters such as the training time and prediction speed as listed in Table \ref{tab:overhead_ML_models} are based on the Matlab cloud server machine. Nevertheless, it is worth mentioning that since we do performance comparison of the ML models relative to each other, the conclusions drawn remain invariant even if the ML models are implemented on another machine with potentially different specifications. Last but not the least, Table \ref{tab:overhead_ML_models} also summarizes the training and test complexity of all the ML classifiers in terms of big-O notation, which is agnostic of the underlying machine executing the ML models. }

\begin{table*}[!htbp]
\caption{Performance benchmarks for ML classifiers with a data augmentation factor of 4, detailing their accuracy, training time, prediction speed, and memory usage. Training and classification complexity of each ML classifier are expressed in Big-O notation, where \( n \) is the number of samples, \( d \) is the number of features, \( t \) is the number of trees in ensemble methods, and \( L \) is the number of layers in the shallow neural network.}

\begin{center}
\begin{tabular}{|l|c|c|c|c|c|c|c|}
\hline
\textbf{Model} & \multicolumn{1}{c|}{\textbf{Accuracy}} & \multicolumn{1}{c|}{\textbf{Training Time}} & \multicolumn{1}{c|}{\textbf{Prediction Speed}}  & \multicolumn{1}{c|}{\textbf{Memory}} & \multicolumn{1}{c|}{\textbf{Training Complexity}} & \multicolumn{1}{c|}{\textbf{Classification Complexity}} \\
 & \textbf{(\%)} & \textbf{(sec)} & \textbf{(obs/sec)}  & \textbf{(MB)} & & \\
\hline
KNN-Euclidean  & 99.9 & 48.034 & 347.28 & 75.44 & $O(1)$ & $O(n * d)$\\
KNN-cosine & 84.9 & 21.259 & 798.25 & 75.44 & $O(1) $&$ O(n * d)$\\
KNN-Cubic  & 89.4 & 484.47 & 32.958 & 75.44 & $O(1)$ & $O(n * d)$\\
Ensemble-SubspaceKNN & 99.9 & 103.22 & 159.88 & 1089.859 & $O(t * n * \log{n} * d)$ & $O(t * n *log(n)* d)$\\
Ensemble-BaggedTree & 99.9 & 92.363 & 34960 & 11.467 & $O(t * n * \log{n} * d)$ &$ O(t * log(n) * d)$\\
Kernel-SVM & 90.2 & 89.68 & 1225.4 & 1.119 & $O(n^2 * d)$ to $O(n^3 * d)$ & $O(n * d)$\\
Kernel-LR & 86.5 & 287.15 & 1273 & 1.119 & $O(n * d)$ &$ O(d)$\\
SVM-Fine Gaussian & 88 & 2.2048 & 3765.6 & 57.29 & $O(n^2 * d$ to $ O(n^3 * d)$ &$ O(n * d)$\\
1x100 NN & 77.3 & 280.65 & 55877 & 0.276 & $O(L * n * d^2)$ & $O(L * d^2)$\\
2x10 NN & 60.8 & 37.002 & 15,526 & 0.068 & $O(L * n * d^2)$ & $O(L * d^2)$\\
3x10 NN & 70.4 & 177.99 & 36857 & 0.070 & $O(L * n * d^2)$ & $O(L * d^2)$\\
\hline
\end{tabular}
\label{tab:overhead_ML_models}
\end{center}
\end{table*}

%%%%%%%

\subsection{Performance Evaluation of Deep Learning and Transformer Models}

%The testing accuracy attained by the lstm model was 87.56\% and the confusion matrix is shown in Figure 5.

\begin{figure*}[htb]
\centering
\subfigure[LSTM-FCN Accuracy]{\includegraphics[width=.24\linewidth]{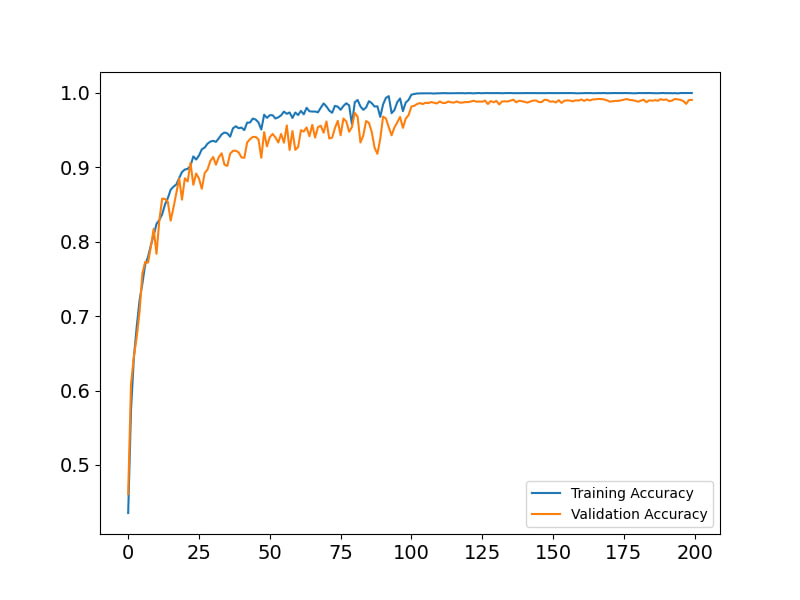}}
\hfill
\subfigure[BiLSTM-FCN Accuracy]{%
  \includegraphics[width=.24\linewidth]{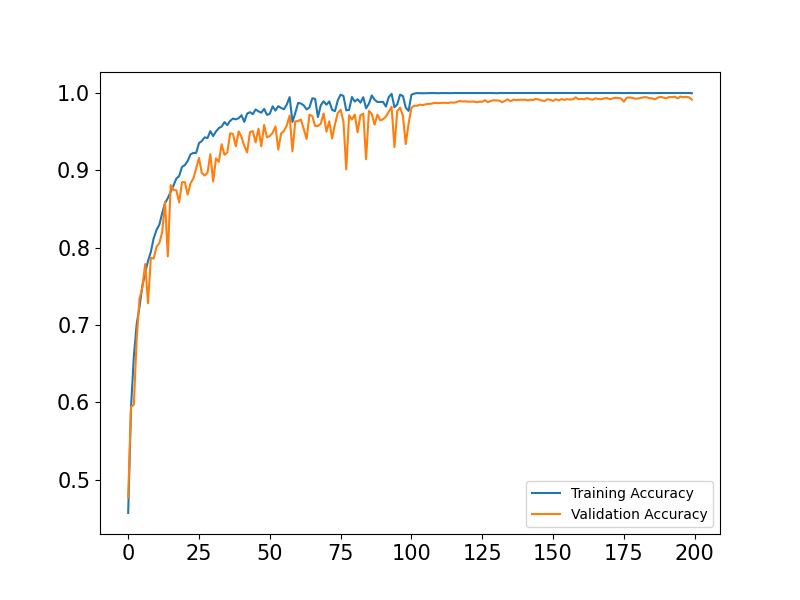}%
}
\hfill % this will fill the space between the figures if needed
\subfigure[DistilBERT Accuracy]{%
  \includegraphics[width=.23\linewidth]{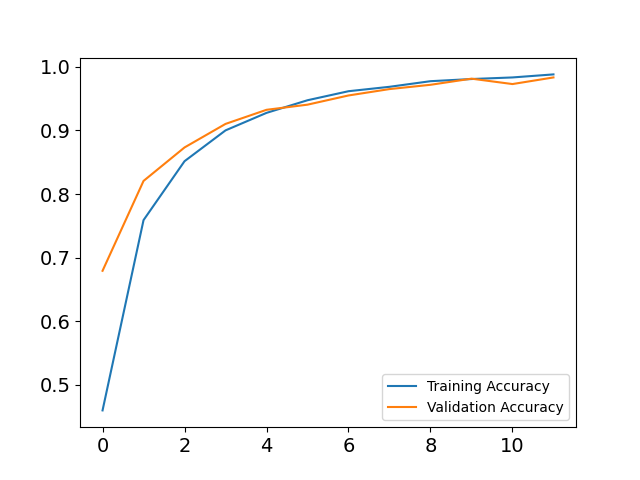}%
}
\hfill % this will fill the space between the figures if needed
\subfigure[1D-ViT Accuracy]{%
  \includegraphics[width=.23\linewidth]{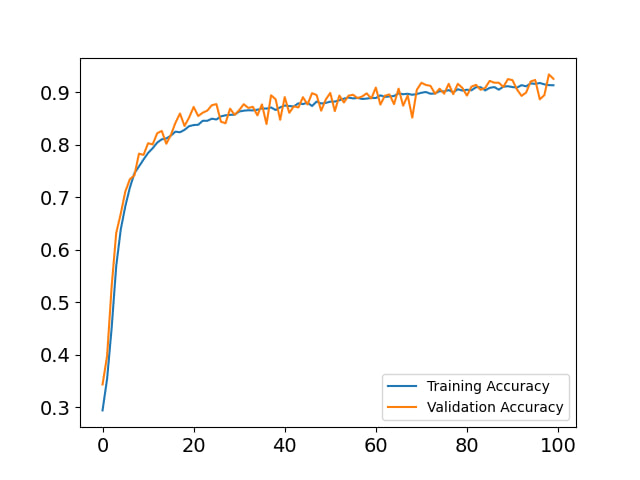}%
}
\\ % this breaks the line for the next row of figures
\subfigure[LSTM-FCN Loss]{%
  \includegraphics[width=.24\linewidth]{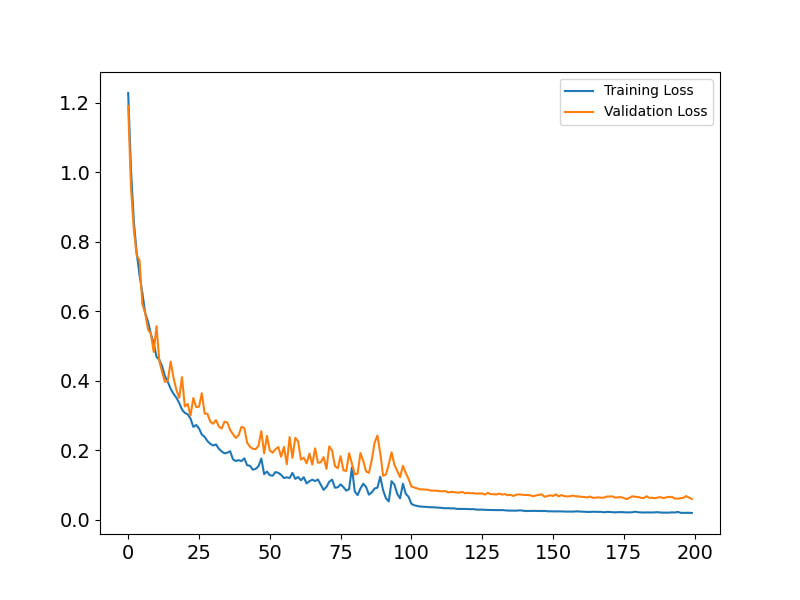}%
}
\hfill % this will fill the space between the figures if needed
\subfigure[BiLSTM-FCN Loss]{%
  \includegraphics[width=.24\linewidth]{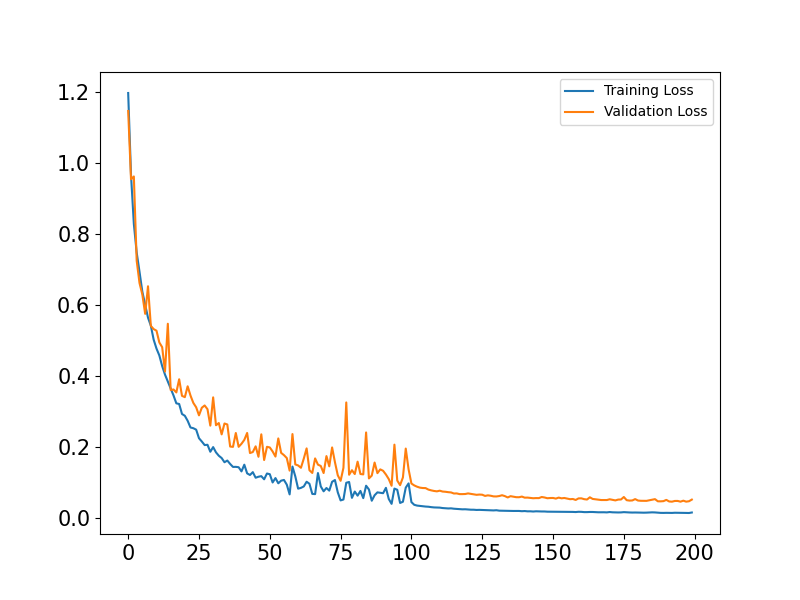}%
}
\hfill 
\subfigure[DistilBERT Loss]{%
  \includegraphics[width=.23\linewidth]{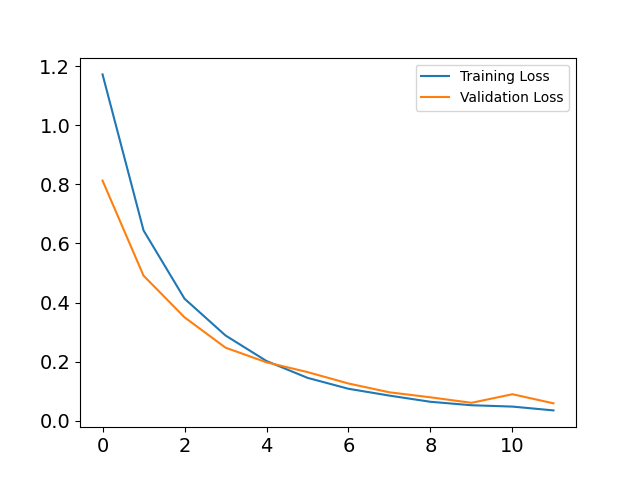}%
}
\hfill % this will fill the space between the figures if needed
\subfigure[1D-ViT Loss]{%
  \includegraphics[width=.23\linewidth]{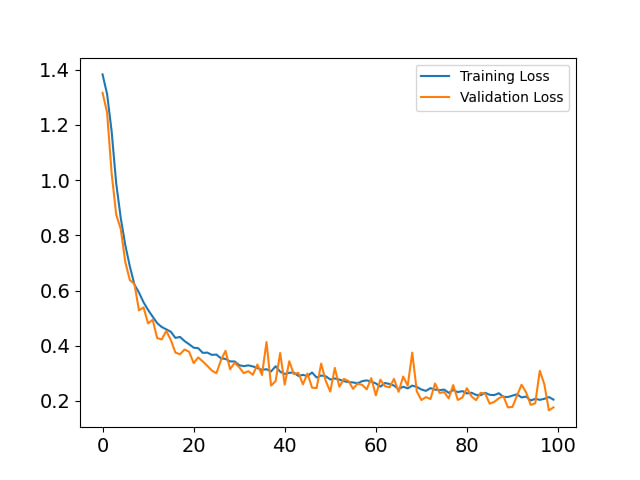}%
}
\hfill 
\caption{Training performance (i.e., Accuracy/Loss vs. epochs) curves for LSTM-FCN, BiLSTM-FCN, DistilBERT, and 1D-ViT models. X-axis represents the number of epochs, while Y-axis represents Accuracy/Loss of a given model. Data augmentation factor is set to 4. }
\label{fig:epochs_plot}
\end{figure*}

We begin with Fig. \ref{fig:epochs_plot} that presents a detailed examination of the training performance of the two DL and two transformer models (for a data augmentation factor of 4). Specifically, for all the four models, we observe a steady increase (decrease) in the accuracy (loss) as we increase the number of epochs in the top (bottom) row of Fig. \ref{fig:epochs_plot}. This is indicative of a robust learning process without any underfitting or overfitting. Especially, the loss and accuracy functions for DistilBERT model reach a plateau in about 10 epochs during the fine-tuning process, which demonstrates the effectiveness of the pre-training of DistilBERT. Similarly, for LSTM-FCN, BiLSTM-FCN, and 1D-ViT models, the accuracy and loss functions saturate in about 100 epochs. Fig. \ref{fig:epochs_plot} also demonstrates the effectiveness of our approach whereby we use a relatively large learning rate in the beginning with the aim of faster convergence of the proposed models, and later, decrease the learning rate in order to make sure that the models reach the plateau in a stable manner without big fluctuations in accuracy. 

\begin{figure*}[htb]
\centering
% \vspace{-4mm}
\subfigure[Accuracy comparison of the deep learning and transformer models for binary classification (hydrated vs. dehydrated) ]{%
  \includegraphics[width=.45\linewidth]{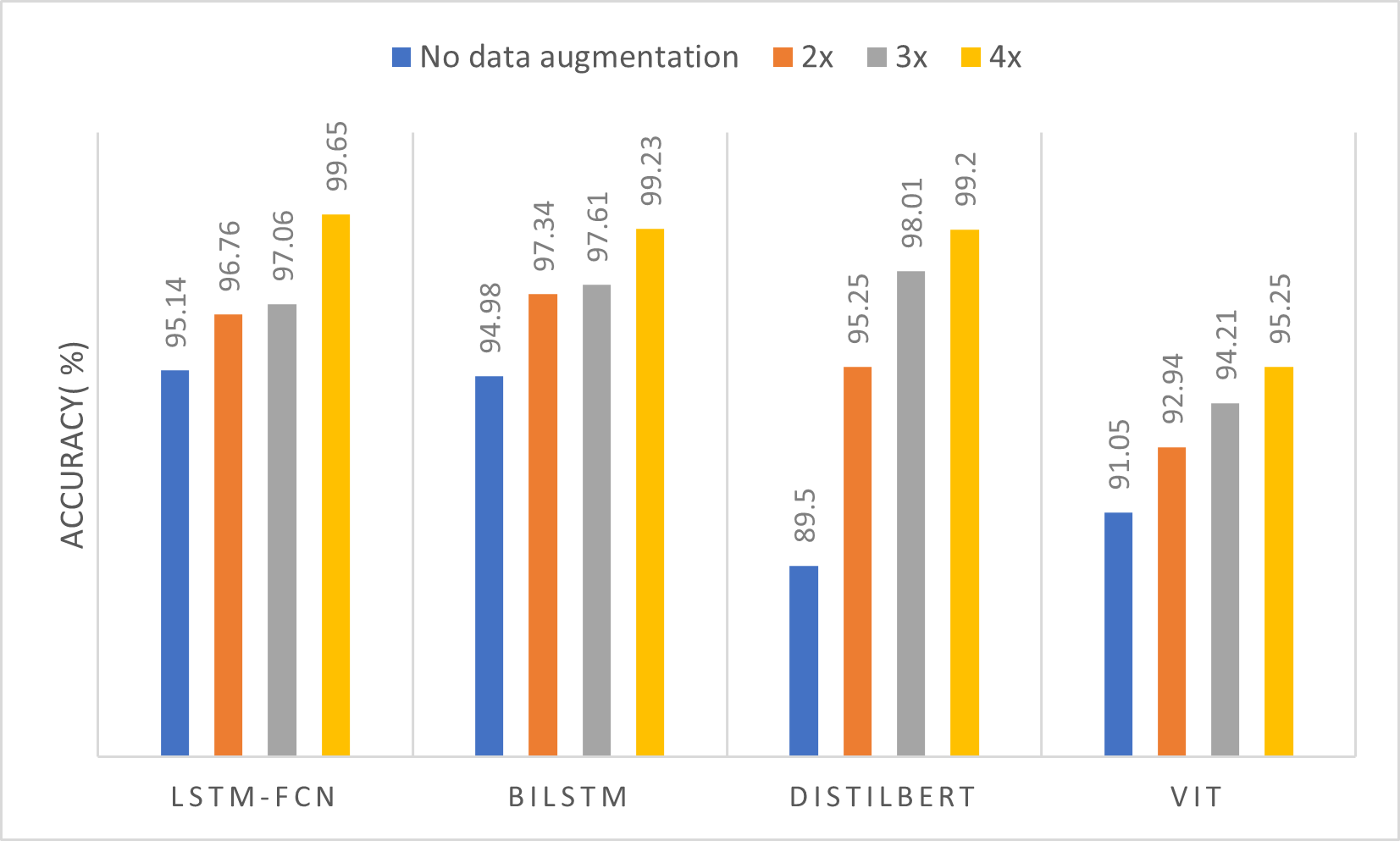}%
}
\hfill
% \vspace{-2mm} % Adjust this value as needed
\subfigure[Accuracy comparison of the deep learning and transformer models for 4-class classification (fully hydrated, mildly dehydrated, moderately dehydrated, extremely dehydrated)]{%
  \includegraphics[width=.45\linewidth]{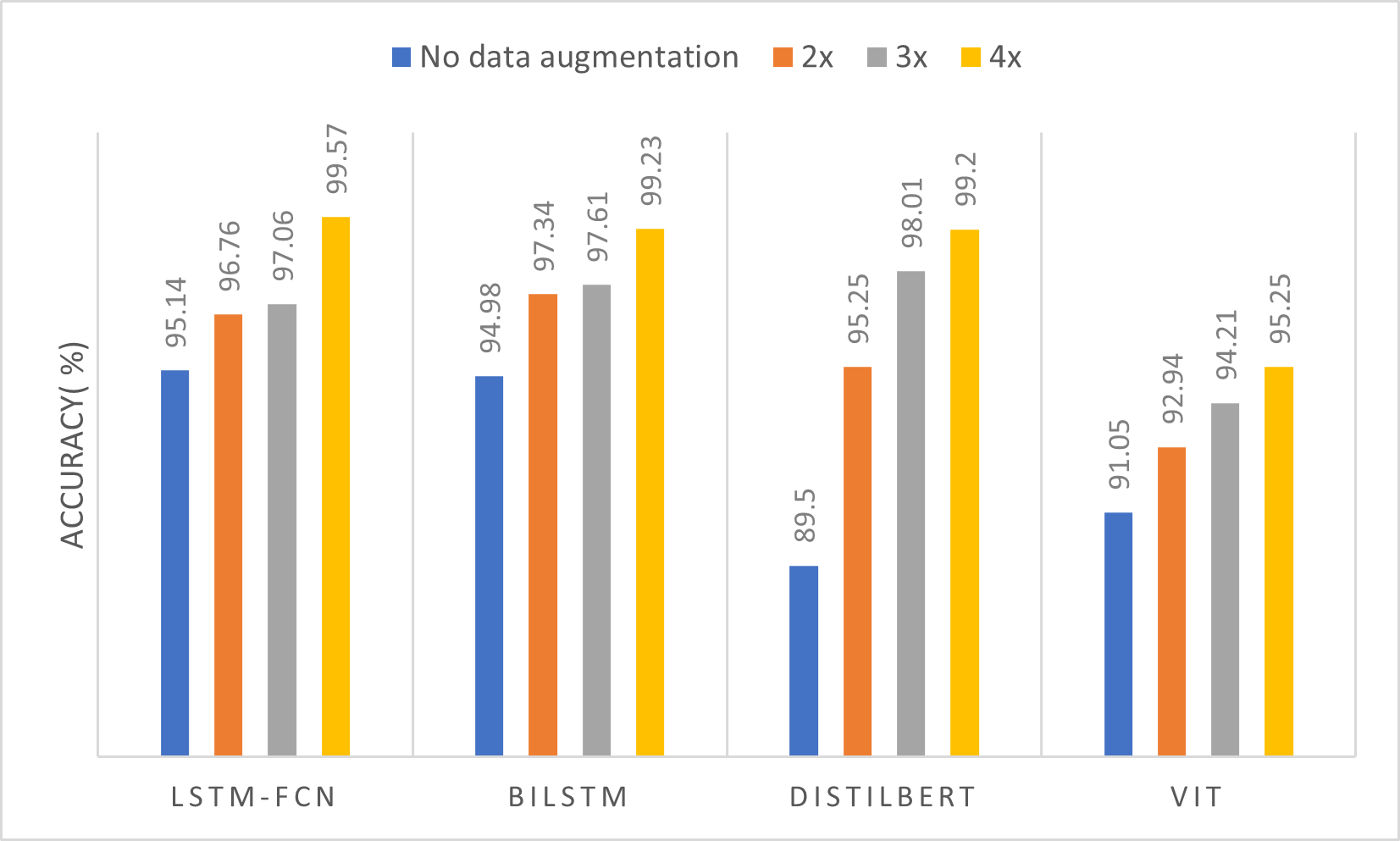}%
}

\caption{Accuracy comparison of LSTM-FCN, BiLSTM-FCN, DistilBERT, and 1D-ViT models for hydration level classification. Sub-fig. (a) compares the accuracy of the four models for binary classification, while sub-fig. (b) compares the accuracy of the four models for 4-class classification. For both sub-figures, we also evaluate the impact of data augmentation (by a factor of 2, 3, and 4) on the accuracy of the four models.
}
\label{fig:accuracy_DL_transformers}
\end{figure*}

Next, Fig. \ref{fig:accuracy_DL_transformers} provides a comprehensive accuracy comparison of LSTM-FCN, BiLSTM-FCN, DistilBERT, and 1D-ViT models for hydration level classification (both binary and 4-class classification) using PPG time-series data. We make the following observations: 
\begin{itemize}
    \item Without any data augmentation, the accuracy of the DL and transformer models is still good as it lies in the range of 89.5\% to 95.14\% (75.67\% to 87.38\%) for binary (4-class) classification. This again points to the feasibility of the problem at hand, i.e., the smartphone-based video-PPG data is ineed a legitimate biomarker of dehydration.
    \item Data augmentation (by factors of 2, 3, 4) leads to a further and monotonic increase in accuracy for all the four models, for both classification problems. For example, for a data augmentation factor of 4, the accuracy of all the four models is very high and lies in a narrow range of 95.25\% to 99.65\% (93.08\% to 99.57\%) for binary (4-class) classification problem. 
    \item For both classification tasks, the data augmentation has more impact on the performance of data-hungry transformer models, compared to DL models. That is, DistilBERT shows dramatic performance improvements with increase in data augmentation factor, thereby indicating the potential need for larger datasets to train more complex models effectively. On the other hand, the LSTM-FCN (and BiLSTM-FCN) model maintains relatively high accuracy in a narrow range, for all data augmentation levels. This reflects the fact that the proposed DL models are well-matched with the PPG time-series data for the given problem of dehydration level classification. 
    \item The performance offset between the accuracy of the lowest-performing model (i.e., 1D-ViT) for binary classification and for the 4-class classification is quite small, i.e., about 2\% (for a data augmentation factor of 4). This is quite impressive keeping in mind that 4-class classification is much more challenging/complex problem.
\end{itemize}
 
\textcolor{red}{
%For binary classification, the DistilBERT model demonstrates a notable trend; as the data augmentation factor increases, there is a consistent improvement in accuracy. This suggests that DistilBERT, with its intricate architecture, benefits significantly from a larger dataset, reinforcing its capability to capture and generalize complex patterns in the data. Particularly, its accuracy jumps from 89.5\% with no data augmentation to 99.2\% with a fourfold augmentation, underlining the effectiveness of data augmentation in enhancing model performance.
}

\begin{figure}[ht]
\begin{center}	\includegraphics[width=9cm,height=5cm]{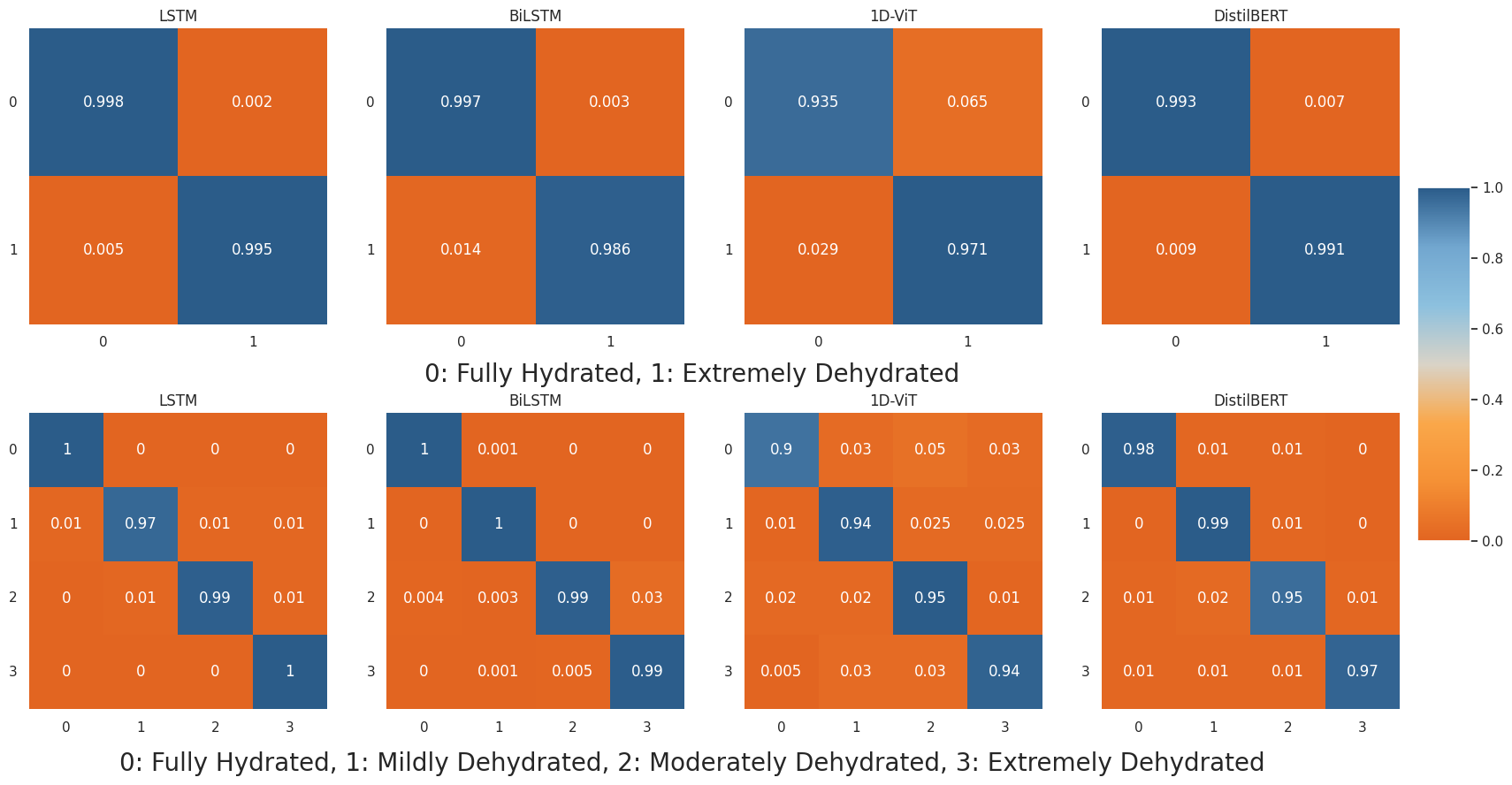} 
\caption{Confusion matrices for LSTM-FCN, BiLSTM-FCN, 1D-ViT, and DistilBERT for dehydration level classification, i.e., binary classification and 4-class classification.}
\label{fig:confusion_matrix}
\end{center}
\end{figure}

Fig. \ref{fig:confusion_matrix} provides the confusion matrices of the DL and transformer models that we implement to do binary (4-class) hydration level classification using the PPG time-series data. Fig. \ref{fig:confusion_matrix} is again an alternate visual illustration of the accuracy results of the DL and transformer models, and thus, strengthens our findings from Fig. \ref{fig:accuracy_DL_transformers}. 
Additionally, Fig. \ref{fig:confusion_matrix} provides more fine-grained details about the performance of each model in terms of sensitivity and specificity across classes, TPRs and FPRs. For example, the TPR of LSTM-FCN model is maximum for the classes 0 and 3, while the TPR of BiLSTM-FCN model is maximum for the classes 0 and 1. 
Note that this result has been generated using a data augmentation factor of 4. 

%Further insights into the predictive precision of these models, including their sensitivity and specificity across classes, are illustrated in the confusion matrices provided in Figure 5. The enhanced dataset not only improved accuracies but also highlighted the trade-offs between computational demand and classification efficacy.

\textcolor{red}{
%the LSTM model reached an impressive accuracy of 98.44\% for binary classification and 95.28\% for the 4-class task. The BiLSTM model closely followed, exhibiting accuracies of 97.80\% for binary and 95.23\% for 4-class classifications. The ViT and DistilBERT models, despite their complexity, showed considerable enhancements in performance with accuracies climbing to 99.20\% for binary and 92.85\% and 90.39\%, respectively, for the 4-class classification when trained with augmented data. This data augmentation notably influenced the advanced models, suggesting their capacity to leverage additional data for improved performance in complex tasks. 
}

Table \ref{tab:DL_transformer_other_benchmarks} provides a quantitative performance comparison of the DL and transformer models in terms of other alternate performance metrics, e.g., number of trainable parameters, memory needs, and FLOPs. 
We observe that the two DL models (LSTM-FCN and BiLSTM models) consists of a few hundred thousands of trainable parameters, which is two orders of magnitude less than the number of trainable parameters of the transformer models (tens of millions of parameters). This difference then translates to FLOPs as well, whereby the FLOPs required by the transformer models are two orders of magnitude greater than the FLOPs required by the DL models. Finally, all the four models require a modest memory storage of a few MBs. All in all, Table \ref{tab:DL_transformer_other_benchmarks} demonstrates that the developed DL models are well-suited for edge computing, i.e., by implementing them as an app on android phones and iPhones. 

\begin{figure}[ht]
\begin{center}	\includegraphics[width=5cm,height=9cm]{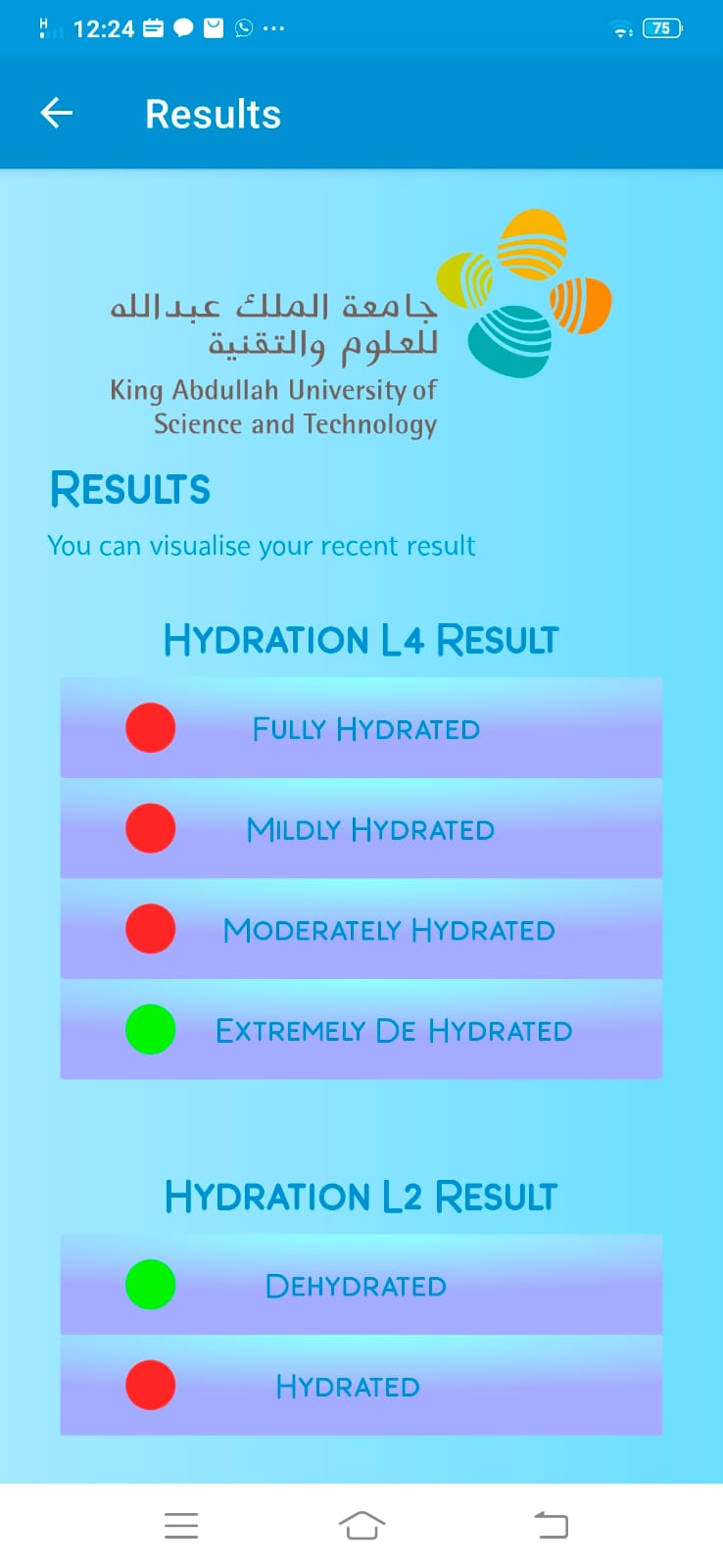} 
\caption{Results page of the custom android app that outputs the dehydration level of a subject on a scale of 1 to 4, based on his/her video-PPG data.}
\label{fig:app}
\end{center}
\end{figure}

In fact, we implement the DL models by means of a custom android app on a smartphone whereby a user places his/her fingertip on the rear camera of his/her smartphone and records a small video of duration 10 seconds. The developed app then pre-processes the raw video data, inputs it to the LSTM-FCN and BiLSTM-FCN models, and subsequently outputs the dehydration level of a subject on a scale of 1 to 4, in about 1 minute. The app is tested to run successfully on an android Vivo phone with the following specs: Vivo model V2024, android version 10, 2 Giga Hertz Snapdragon 665 octa-core processor, 4 GB RAM, and 128 GB storage. For the sake of illustration, Fig. \ref{fig:app} displays the results page of the custom android app. It is worth mentioning that we are using our custom android app to validate and test the DL models on the new, unseen data nowadays. We also plan to add a feature in the android app whereby it will raise an alert and provide the phone user with the suggestions for rehydration if sufficient decrease in hydration level is detected by the app. 

%At this point, it is also worth mentioning that researchers have recently implemented an LLM successfully on an regular (not high-end) android phone \cite{}. 

%The LSTM and BiLSTM models maintained their computational efficiency, as indicated in Table A, with lower memory usage and FLOPs in comparison to the more resource-intensive ViT and DistilBERT models. 

% \begin{figure*}
%     \centering
%     \subfigure[]{\includegraphics[width=0.24\textwidth]{lstm_4_ac.jpg}} 
%     \subfigure[]{\includegraphics[width=0.24\textwidth]{lstm_4_ac.jpg}} 
%     % \subfigure[]{\includegraphics[width=0.24\textwidth]{monalisa.jpg}}
%     % \subfigure[]{\includegraphics[width=0.24\textwidth]{monalisa.jpg}}
%     \caption{(a) blah (b) blah (c) blah (d) blah}
%     \label{fig:foobar}
% \end{figure*}

\begin{table}[htbp]
\caption{Performance comparison of DL and transformer models: other benchmarks (for a data augmentation factor of 4)}
\begin{center}
\begin{tabular}{|p{1.7cm}|p{1cm}|p{1.5cm}|p{1cm}|p{1.5cm}|}
\hline
\textbf{Model} & \textbf{Accuracy (\%)} & \textbf{Number of Parameters} & \textbf{Memory (MB)} & \textbf{FLOPs} \\
\hline
\multicolumn{5}{|c|}{Binary classification} \\
\hline
LSTM-FCN & 99.65& 397,154 & 4.650 & 371,817 \\
BiLSTM-FCN & 99.23 & 414,690 & 5.704 & 531,215 \\
1D-ViT & 95.25 & 22,136,580 & 2.895 & 17,899,821 \\
DistilBERT & 99.20 & 66,363,649 & 8.523 & 236,615,078 \\
\hline
\multicolumn{5}{|c|}{4-class classification} \\
\hline
LSTM-FCN & 99.13& 397,348 & 4.65 & 475,405 \\
BiLSTM-FCN & 99.57 & 415,012 & 5.705 & 513,530 \\
1D-ViT & 93.08 & 18,004,356 & 2.811 & 22,025,517 \\
DistilBERT & 97.42 & 66,365,956 & 8.466 & 236,515,198\\
\hline
\end{tabular}
\label{tab:DL_transformer_other_benchmarks}
\end{center}
\end{table}

\subsection{Performance comparison with related work}

Table \ref{table:comparewithsota} provides a compact but comprehensive performance comparison of our work with the state-of-the-art on non-invasive dehydration monitoring. Our work stands out in terms of the following:
\begin{itemize}
    \item The previous works have utilized various sensing modalities and various sensors, e.g., PPG-sensor-based oximeters, electrodermal activity (EDA)-sensor-based smart watches, bio-chemical analysis of sweat, computer vision-based analysis of the RGB camera-based videos of the face, etc., for non-invasive dehydration monitoring. In contrast, {\it our work utilizes video-PPG data acquired from a smartphone camera for non-invasive dehydration monitoring, for the first time.}
    \item All previous works do dehydration monitoring by means of binary or tertiary (3-class) classification only. In contrast, {\it our work not only does the binary classification, but it could also efficiently provide the dehydration level of a person on a scale of 1 to 4.}
    \item The maximum reported accuracy in the literature for binary classification problem is 97.38\% (by Liaqat et al. \cite{liaqat2022personalized}). On the other hand, a number of the developed ML, DL and transformer models provide an accuracy of more than 99\%. Thus, {\it our ML, DL and transformer models outperform all the methods previously reported in the literature. Additionally, the proposed models report a remarkable maximum accuracy of 99.57\% for the 4-class classification problem, for the first time. }
\end{itemize}
Finally, thanks to the mass proliferation of the smartphones in the society, the proposed method is anticipated to find its application in many diverse usecase scenarios, e.g., self-testing by athletes, marathon runners, labor people working outdoor in the sun, chronic stay-at-home patients, etc. 

\begin{table}[t!]
\centering
\caption{Accuracy comparison of the proposed smartphone-based method for non-invasive dehydration monitoring with the state-of-the-art. }
\label{table:comparewithsota}
\begin{tabular}{|c| c| c| c|} 
 \hline
 {\bf Work} & {\bf Method} & {\bf Modality} & {\bf Accuracy}
\\
\hline
Liaqat et al. \cite{liaqat2022personalized} & Non-invasive & EDA & 97.83$\%$ \\
\hline
Kulkarni et al. \cite{kulkarni2021non} & Non-invasive & EDA & 75.96$\%$ \\
\hline
Liaqat et al. \cite{liaqat2020non} & Non-invasive & EDA & 91.53$\%$ \\
\hline 
Rizwan et al \cite{rizwan2020non} & Non-invasive & EDA & 87.78%$\%$ 
\\
\hline 
% Carrieri et al. \cite{carrieri2020explainable} & Non-invasive & ? & 73.91 $\%$ \\
% \hline  
Saha et al.  \cite{sm13} & Non-invasive & Facial images & 76.1$\%$ \\
\hline
Mengistu et. al.  \cite{7759295} & Non-invasive & acoustic& 91.5$\%$ \\
\hline
Reljin et. al.  \cite{reljin2018automatic} & Non-invasive & PPG& 67.91$\%$ \\
\hline
Hasan et. al. (CBDM)  \cite{hasanhydrationpaper}& Non-contact & RF & 93.8$\%$ \\
\hline 
Hasan et. al. (HBDM)  \cite{hasanhydrationpaper} & Non-contact & RF & 96.15$\%$ \\
\hline

\cellcolor{green!25}This work (2 classes) & \cellcolor{green!25}Non-invasive & \cellcolor{green!25}video-PPG & \cellcolor{green!25}99.65$\%$ \\
\hline
\cellcolor{green!25}This work (4 classes) & \cellcolor{green!25}Non-invasive & \cellcolor{green!25}video-PPG & \cellcolor{green!25}99.65$\%$ \\
\hline 

\end{tabular}
\vspace{-1mm}
\end{table}

 % \begin{figure}[ht]
 % \begin{center}
% 	\includegraphics[width=9cm,height=5cm]{Figures/TL2.png} 
% \caption{Dehydration classification based on Transfer learning.}
% \label{fig:sysmodel}
% \end{center}
% \end{figure}

\section{Data Visualization, An Alternate Low-Complexity Method for Dehydration Monitoring and Interpretation of Model Predictions}

Having presented the results that showcase the very high accuracy obtained by the proposed ML, DL and transformer models for dehydration level classification, we now turn our attention to the following: 
\begin{itemize}
    \item {\it Data Visualization:} We do PPG data visualization in a lower-dimensional (2D/3D) feature space by means of a well-known dimensionality reduction method known as t-distributed stochastic neighbor embedding (t-SNE). We observe that the raw PPG data when visualized through t-SNE leads to the two or four classes being entangled. We resolve this issue by passing the PPG data through an LSTM-FCN model to the t-SNE. This in turn prompts us to design an alternate low-complexity method for dehydration monitoring as follows.
    \item {\it An Alternate Low-Complexity Method for Dehydration Monitoring:} We first feed the raw PPG data to an LSTM-FCN model (basically, a variant of the model described in Fig. \ref{fig:diagram_lstm_bilstm}) that does feature extraction. The extracted features are then passed through t-SNE algorithm that does dimensionality reduction (or, feature selection). Finally, the low-dimensional (2D/3D) data is fed to a number of ML methods which efficiently do dehydration level classification. 
    \item {\it Interpretation of Model Predictions:} We provide interpretation of predictions by our best-performing LSTM-FCN model, under the explainable AI (XAI) framework. With this, we aim to not only validate the discriminative power of the features extracted by the developed DL model, but also to enhance the transparency and trustworthiness of the predictions by the DL model. 
\end{itemize}

\subsection{PPG Data Visualization by Dimensionality Reduction via t-SNE Method}

\begin{figure*}[htb]
\centering

\subfigure[2D visualization without using LSTM-FCN  (Binary classification)]{%
  \includegraphics[width=.25\linewidth]{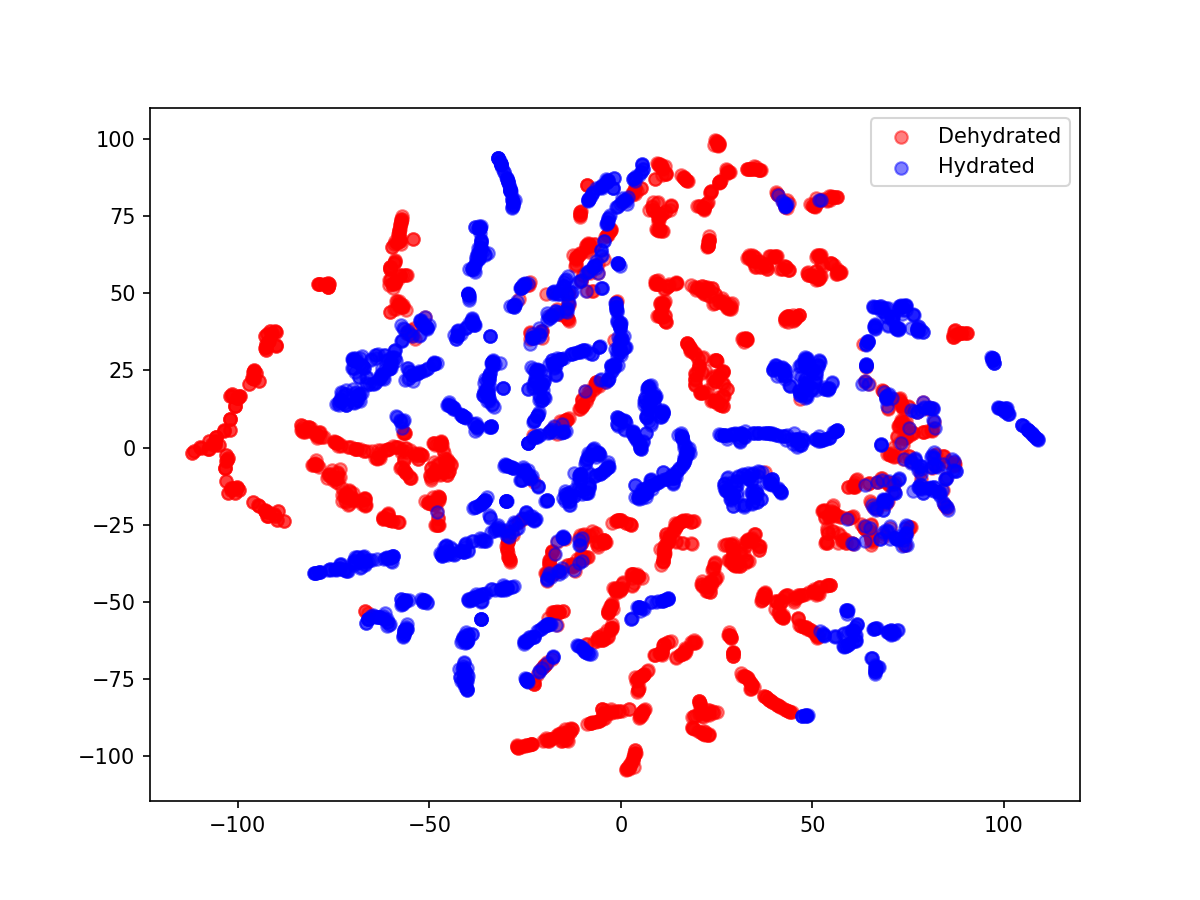}%
}
\hfill
\subfigure[2D visualization using LSTM-FCN  (Binary classification)]{
  \includegraphics[width=.24\linewidth]{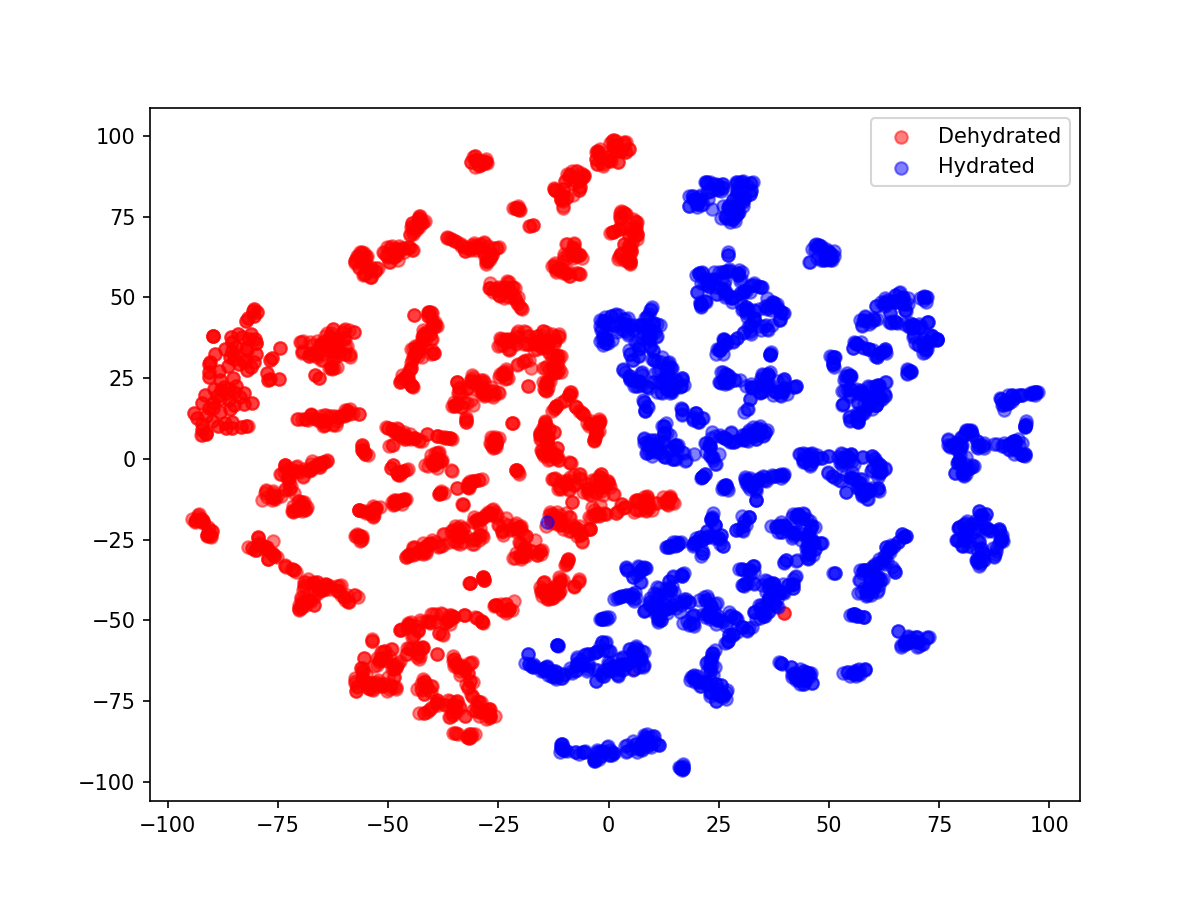}}
\hfill
\subfigure[3D visualization without using LSTM-FCN  (Binary classification)]{%
  \includegraphics[width=.20\linewidth]{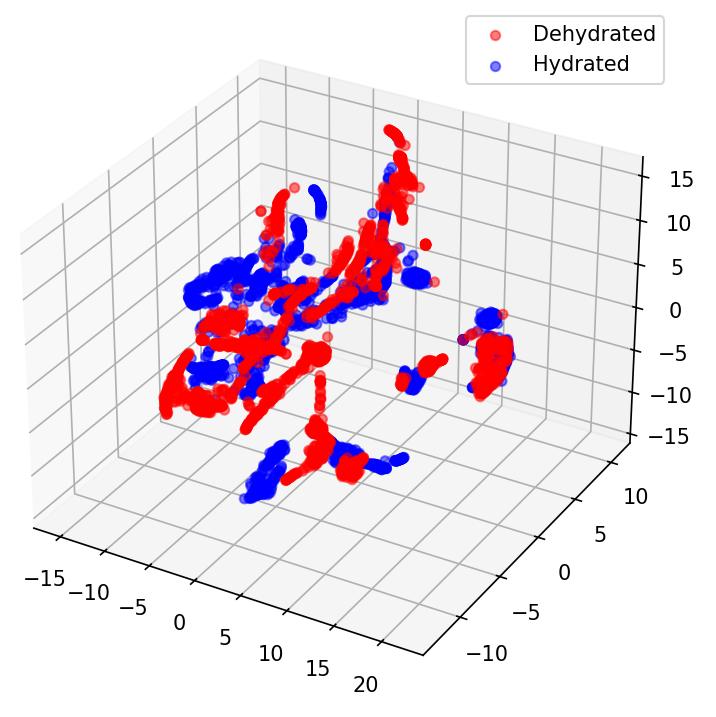}%
}
\hfill % this will fill the space between the figures if needed
\subfigure[3D visualization using LSTM-FCN (Binary classification)]{%
  \includegraphics[width=.21\linewidth]{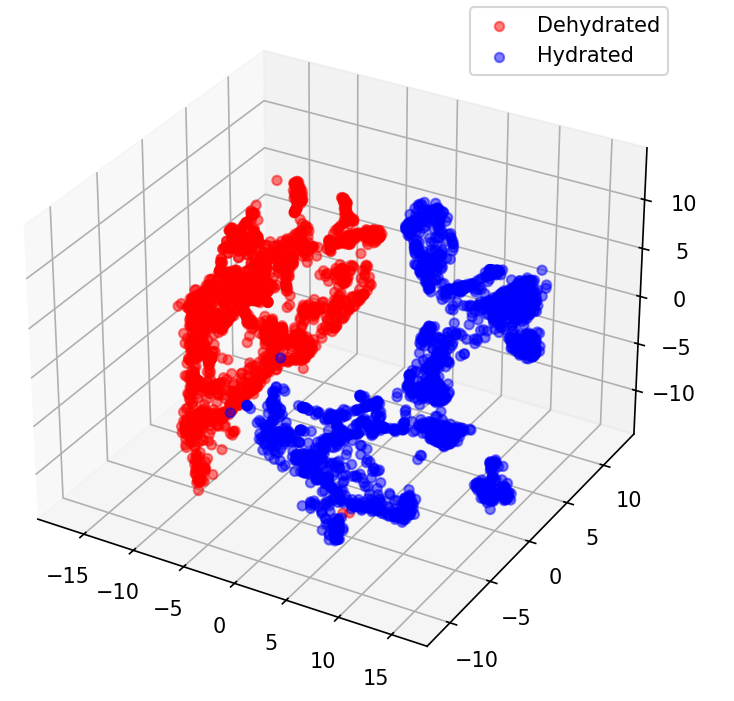}%
}
\\ % this breaks the line for the next row of figures

\subfigure[2D visualization without using LSTM-FCN (4-class classification)]{%
  \includegraphics[width=.26\linewidth]{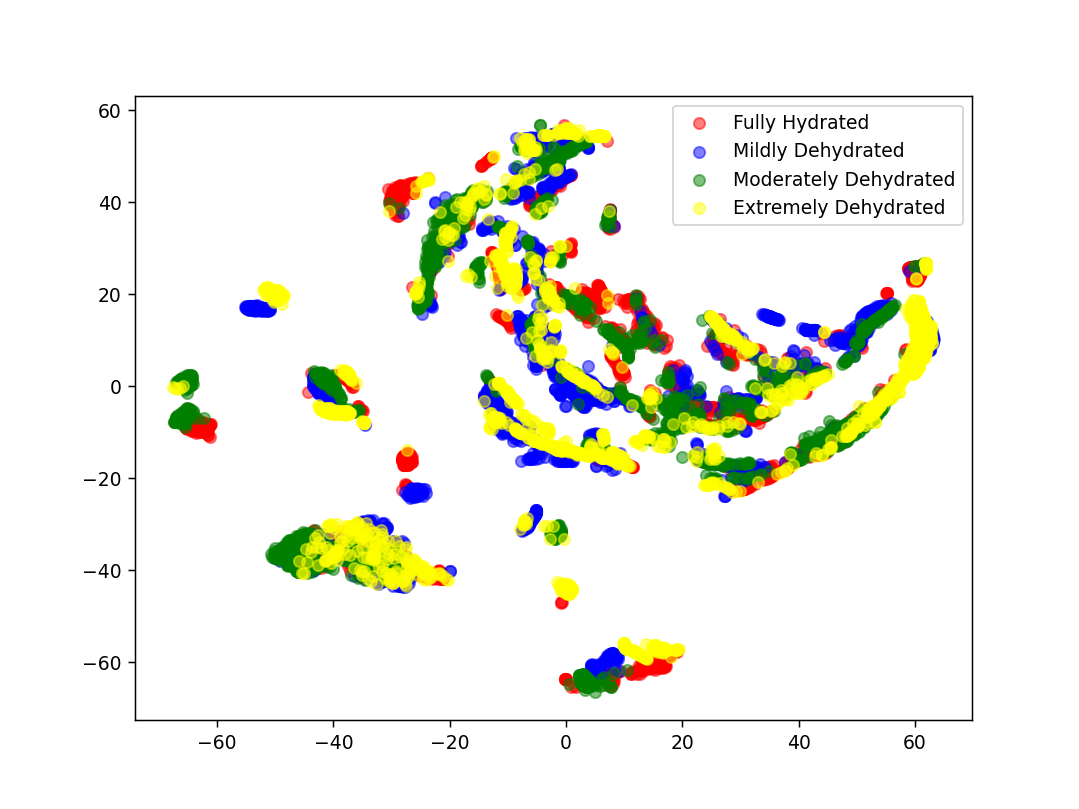}%
}
\hfill
\subfigure[2D visualization using LSTM-FCN  (4-class classification)]{%
  \includegraphics[width=.26\linewidth]{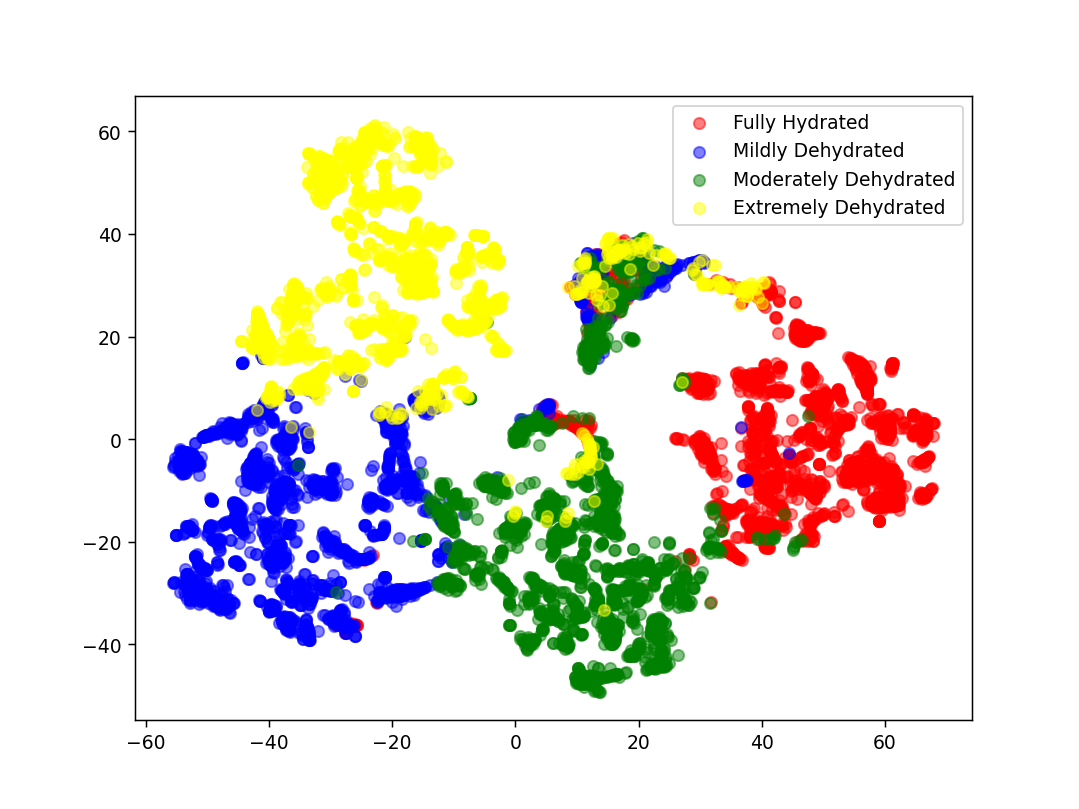}%
}
\hfill
\subfigure[3D visualization without using LSTM-FCN  (4-class classification)]{%
  \includegraphics[width=.20\linewidth]{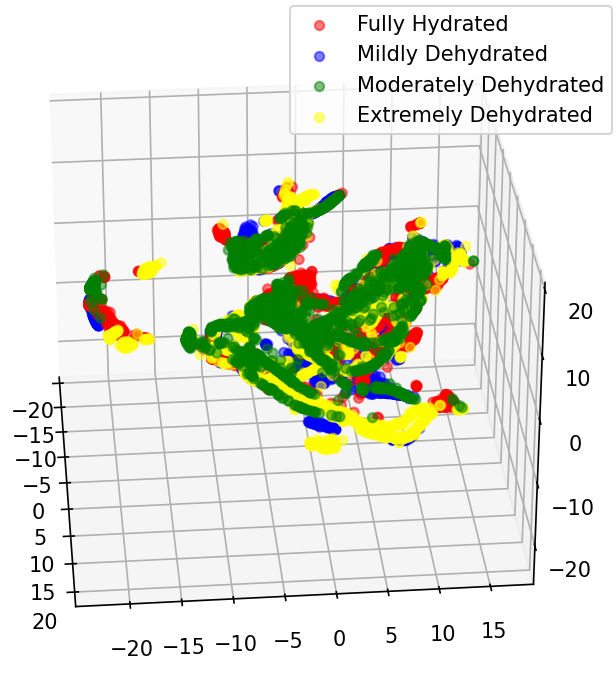}%
}
\hfill % this will fill the space between the figures if needed
\subfigure[3D visualization using LSTM-FCN (4-class classification)]{%
  \includegraphics[width=.20\linewidth]{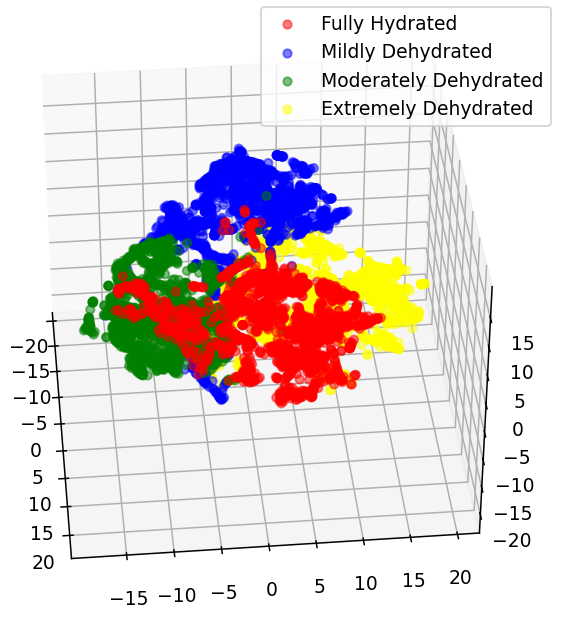}%
}
\caption{2D and 3D t-SNE visualization of PPG time-series data, with and without using LSTM-FCN Features.}
\label{fig:tsne}
\end{figure*}

We utilize the well-known t-SNE algorithm that projects the high-dimensional PPG data into a low-dimensional (2D/3D) feature space. Specifically, we feed the raw PPG segments, each spanning 3.3 seconds (or, having 100 samples), to the t-SNE method in order to construct a visual map of data distribution in 2D/3D space. Fig. \ref{fig:tsne} (a), (c) show the resulting t-SNE visualizations of the data for the binary classification problem in 2D, 3D space, respectively. Similarly, Fig. \ref{fig:tsne} (e), (g) show the resulting t-SNE visualizations of the data for the 4-class classification problem in 2D, 3D space, respectively. We observe from the Fig. \ref{fig:tsne} (a),(c),(e),(g) that the data points corresponding to different classes in the dataset are entangled. This points to the inherent challenge in distinguishing between classes, with data points appearing intermixed within the feature space, emphasizing the intricate nature of PPG data. This prompts us to feed the PPG segments through an LSTM-FCN model (basically, the same model as described in Fig. \ref{fig:diagram_lstm_bilstm}, except the softmax layer), whose output is then fed to the t-SNE algorithm. Fig. \ref{fig:tsne} (b), (d) show the resulting t-SNE visualizations of the feature vector for the binary classification problem in 2D, 3D space, respectively. Similarly, Fig. \ref{fig:tsne} (f), (h) show the resulting t-SNE visualizations of the feature vector for the 4-class classification problem in 2D, 3D space, respectively. We note that the data points corresponding to different classes are more discernibly separated now, which points to the enhanced feature extraction capability of the developed LSTM-FCN model. Thus, LSTM-FCN and t-SNE together facilitate the distinction between hydrated and dehydrated states, as well as the gradation of hydration levels on a four-tier scale. Additionally, the t-SNE method reduces the size of the feature vector from $100\times 1$ to $g\times 1$, with $g$ being 2 or 3. This in turn motivates us to design an alternate low-complexity method for dehydration monitoring which we discuss in the next sub-section.

%Together, these visualizations validate the strength of LSTM-FCN model in particular and DL models in general, for feature extraction. 

%To effectively illustrate the distribution of our PPG signal data in feature space, t-distributed Stochastic Neighbor Embedding (t-SNE) was employed for effective visualization. 
%  \begin{figure}[H]
% \CommonHeightRow{%
% \begin{floatrow}[2]%
%     \ffigbox[\FBwidth]
%     {\includegraphics[height=\CommonHeight]{Figures/TSNE_2_1.png}
%     {\caption*{Energy Vs Photon Count graph for day 44.6}}
%     \ffigbox[\FBwidth]
%  {\includegraphics[height=\CommonHeight]{Figures/TSNE_2_2.png}}
%     {\caption*{Energy Vs Photon Count graph for day 45.9}}
% \end{floatrow}}%
%     \end{figure}

\subsection{A LSTM-FCN-plus-t-SNE based low-complexity method for dehydration level classification}

\begin{figure}[ht]
\begin{center}	\includegraphics[width=3.5cm,height=5cm]{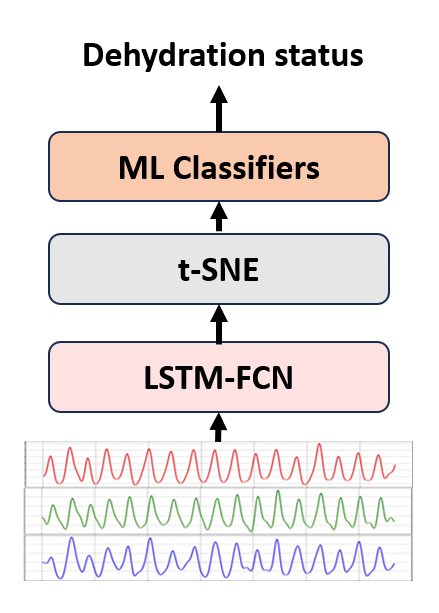} 
\caption{ Block diagram of an alternate low-complexity method, that utilizes LSTM-FCN model for feature extraction from the PPG time-series data, utilizes the t-SNE method for feature selection, and implements a number of ML classifiers for dehydration level classification. }
\label{fig:tsne_classification_method}
\end{center}
\end{figure}

We now discuss the particulars of an alternate low-complexity method, that utilizes the LSTM-FCN model for feature extraction from the PPG time-series data, utilizes the t-SNE method for feature selection, and implements a number of ML classifiers for dehydration level classification (see Fig. \ref{fig:tsne_classification_method}). 
Recall from Fig. \ref{fig:tsne} that the visibly segregated clusters formed due to feeding of the PPG data to LSTM-FCN followed by t-SNE serve as a foundation for the hydration level classification. Subsequently, our alternate method employs a number of ML classifiers in order to interpret and differentiate these clusters corresponding to different classes in order to make predictions about the hydration status of a person. Note that the performance of the ML classifiers is greatly dependent on the quality of the feature extraction process (by LSTM-FCN) as well as the feature selection process (by t-SNE). Our workflow in Fig. \ref{fig:tsne_classification_method}, therefore, emphasizes the synergy between feature extraction, dimensionality reduction, and classification, leading to an effective methodology for predicting dehydration status from PPG time-series data.

In Table \ref{tab:tsne_model_accuracies}, we systematically compile the accuracy results of our alternate method, which in turn relies upon the proficiency of various ML classifiers in differentiating between various hydration states. For the binary classification problem, the accuracy of all the ML models (except Kernal-LR and Kernal-SVM) is very high and lies in a very narrow range of 89.8\% to 96.8\%, 94.7\% to 98.6\% for 2D, 3D case, respectively. Similarly, for the 4-class classification problem, the accuracy of all the ML models (except Ensemble-subspace KNN) is again very high and lies in a very narrow range of 92.2\% to 93.9\%, 90.2\% to 94.1\% for 2D, 3D case, respectively. Essentially, Table \ref{tab:tsne_model_accuracies} reflects the effectiveness of the purpose-built alternate method that feeds high-dimensional PPG time-series data to LSTM-FCN model for feature extraction, followed by t-SNE for feature selection and dimensionality reduction, followed by a number of ML classifiers that utilize a feature vector of very small size (2 or 3) in order to do dehydration level classification. 

%In the realm of binary classification, the KNN-Euclidean model with $k=10$ demonstrated exceptional accuracy, achieving highest accuracy of 98.6\% for the 3D case (i.e., using a feature vector of size 3 only). This model also showed formidable performance in the more complex problem of 4-class classification, achieving an accuracy of 94.2\%. The shallow neural network (1x100 NN model), on the other hand, demonstrated notable performance for the 2D case (i.e., feature vector of size 2) for binary classification, achieving an accuracy of 95.8\%. 

\begin{table}[H]
\centering
\caption{Accuracy of the ML Models under the proposed alternate LSTM-FCN-plus-t-SNE based low-complexity method for
dehydration level classification.}
\label{tab:tsne_model_accuracies}
\small % Make the font size smaller to fit the table within the column
\begin{tabular}{lcccc}
\hline
\textbf{Model} & \multicolumn{4}{c}{\textbf{Accuracy (\%)}} \\ 
& \multicolumn{2}{c}{\textbf{Binary}} & \multicolumn{2}{c}{\textbf{4-Class}} \\ 
& \textbf{2D} & \textbf{3D} & \textbf{2D} & \textbf{3D} \\ \hline

KNN-Euclidean & 95.8 & 96.5 & 92.2 & 93.6 \\
% KNN-Eucl k=10 & 96.8\% & 98.6\% & 93.5\% & 94.2\% \\
% KNN-Coarse k=100 & 95.4\% & 95.4\% & 92.4\% & 93.2\% \\
KNN-Cosine & 92.1 & 94.9 & 92.6 & 90.2 \\
KNN-Cubic & 96.8 & 98.6 & 93.5 & 93.9 \\
Ensemble - Subspace KNN & 89.8 & 94.7 & 61.7 & 91.1 \\
Ensemble-Bagged Trees & 96.3 & 96.8 & 93.4 & 93.5 \\
Kernal-SVM & 92.1 & 83.8 & 92.6 & 93.6 \\
Kernal-LR & 92.1 & 76.6 & 92.6 & 92.9 \\
SVM-Fine Gaussian  & 95.6 & 97.9 & 93.9 & 93.5 \\
2x10 NN & 95.1 & 96.3 & 93.4 & 93.3 \\
1x100 NN & 95.8 & 98.1 & 93.5 & 94.1 \\
3x10 NN & 95.4 & 95.8 & 92.4 & 93.2 \\
% 1x25 NN & 94.9\% & 97.9\% & 93.5\% & 93.5\% \\
\hline
\end{tabular}
\end{table}

\subsection{XAI: SHAP Analysis of the Designed LSTM-FCN Model for Non-Invasive Dehydration Monitoring}

Recall that we implement more than a dozen AI models for dehydration level classification, and a majority of the developed models reported a very high accuracy in a narrow range of 95\% to 99\% (for both classification problems), which is quite satisfactory. Nevertheless, it is imperative to look for insights into the internal decision-making process of these (block-box) classifiers that do dehydration level classification using PPG time-series data analysis. Thus, to demystify these models and elucidate the key features driving their predictions, we employ XAI techniques that aim to shed light on the opaque nature of decision-making process of the AI models. The XAI approach not only enhances the transparency of the designed AI models, but also empowers us with a deeper understanding of the physiological indicators of the hydration level of a person. 

Having said that, we utilize SHAP (SHapley Additive exPlanations) method \cite{lundberg2017unified} with the aim to highlight the most influential features within the PPG segments that contribute to the decision-making by the proposed AI models. More precisely, we employ SHAP Gradient Explainer, a feature attribution technique that is based on integrated gradients method \cite{shapxai} to elucidate the predictions made by our best-performing LSTM-FCN model using the RGB PPG time-series data. By leveraging this approach, we compute a SHAP value for every input feature within the PPG segment across all three color channels. From this set of SHAP values, we identify and highlight the top 10 SHAP values per PPG color channel, which represent the most influential features that contribute to the diagnostic interpretation of a specific PPG segment.

\begin{figure*}[htb]
\centering

\subfigure[]{%
  \includegraphics[width=.22\linewidth]{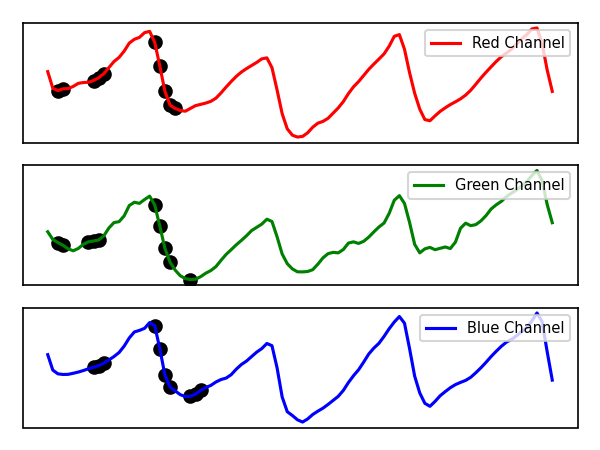}%
}
\hfill
\centering
\subfigure[]{%
  \includegraphics[width=.22\linewidth]{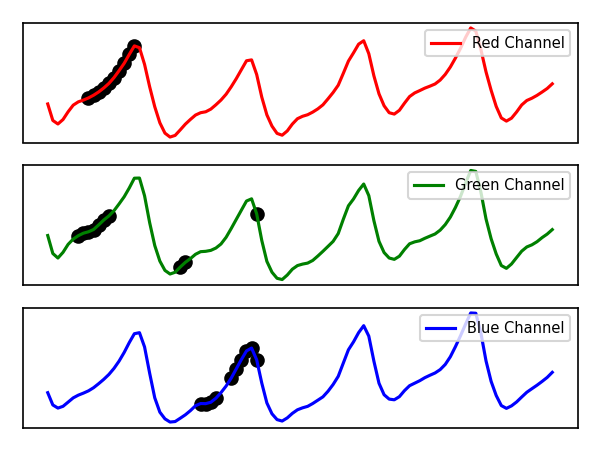}%
}
\hfill % this will fill the space between the figures if needed
\centering
\subfigure[]{%
  \includegraphics[width=.22\linewidth]{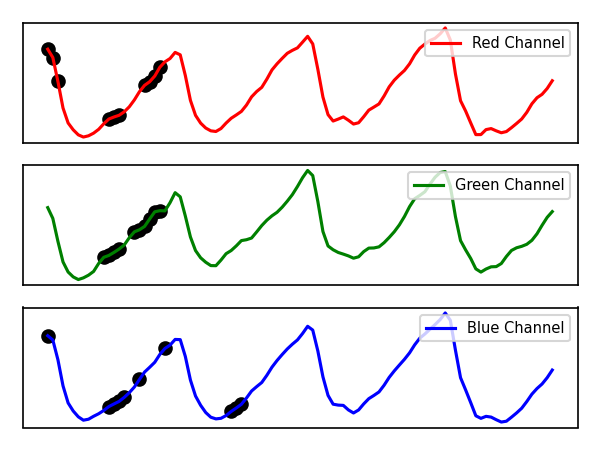}%
}
\hfill % this will fill the space between the figures if needed
\centering
\subfigure[]{%
  \includegraphics[width=.22\linewidth]{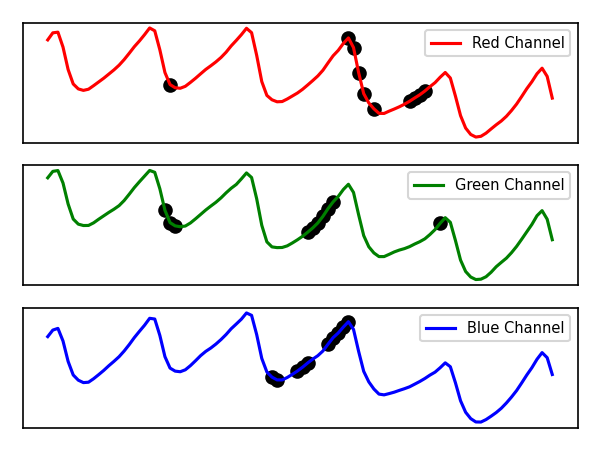}%
}
\\ % this breaks the line for the next row of figures
\centering
\subfigure[]{%
  \includegraphics[width=.22\linewidth]{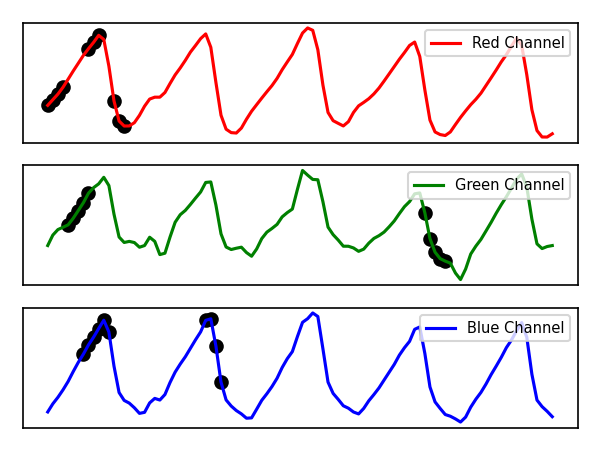}%
}
\hfill % this will fill the space between the figures if needed
\centering
\subfigure[]{%
  \includegraphics[width=.22\linewidth]{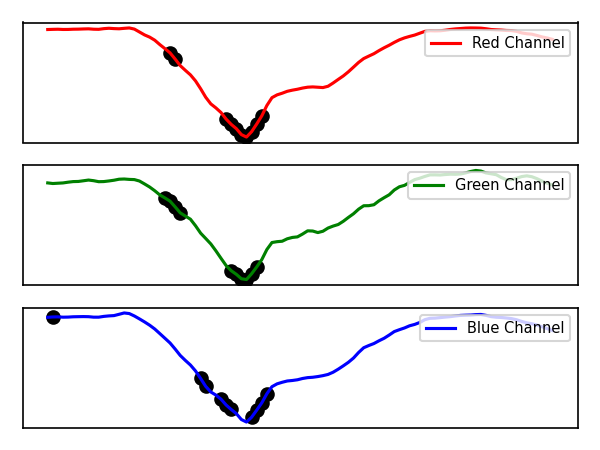}%
}
\hfill
\centering
\subfigure[]{%
  \includegraphics[width=.22\linewidth]{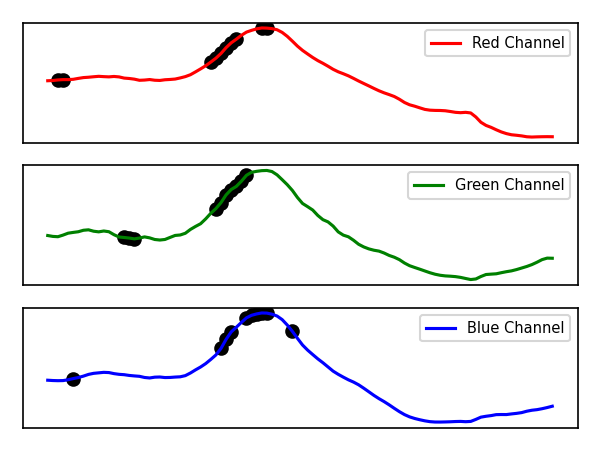}%
}
\hfill % this will fill the space between the figures if needed
\subfigure[]{%
  \includegraphics[width=.22\linewidth]{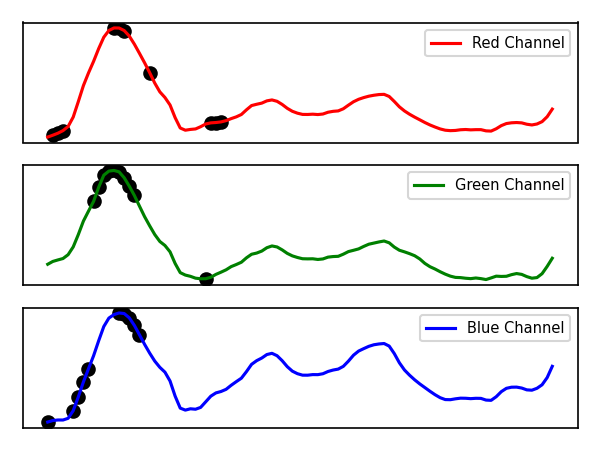}%
}
\hfill 
\caption{SHAP method-based interpretation of the decision-making by LSTM-FCN model by identification of 10 most significant SHAP features (see black dots) in RGB PPG segments of length 100 samples each. Sub-fig. (a)-(d) highlight SHAP-determined features pinpointing critical areas influencing model predictions, for clean PPG segments. Sub-fig. (e) shows a situation where significant SHAP features for the three PPG color channels span across multiple cardiac cycles. Sub-fig. (f)-(h) do SHAP-analysis for low-quality PPG segments that contain distortions or artifacts.}
\label{fig:shap}
\end{figure*}

Fig. \ref{fig:shap} provides a visual depiction of SHAP-based XAI analysis of the LSTM-FCN model, for eight representative PPG segments (each consisting of three color channels). Specifically, SHAP method takes as input an RGB PPG segment of length 3.3 seconds (or, 100 samples), and eventually highlights the top 10 significant features (i.e., samples) contributing to the LSTM-FCN model's decision-making, by marking them with black dots. These features, identified through SHAP values, help us identify the pivotal regions within the PPG segment that are critical for classification. We learn from Fig. \ref{fig:shap} (a)-(d) that the most significant features identified by the SHAP method represent those samples of PPG segments which correspond to: 1) the decay of the PPG waveform (that may occur during the diastole period), 2) rise in the PPG waveform (that corresponds to the systole period), and 3) the moments of noticeable change in the amplitude of the PPG waveform (e.g., systolic peak, etc.). This implies that the morphology of the PPG waveform (i.e., the rate of change in amplitude and more), and thus, the underlying physiology is captured by the LSTM-FCN and SHAP models, which then plays a key role in informing the model's decision-making process. Further, Fig. \ref{fig:shap} (e) captures a rather different situation where the significant SHAP features for the three PPG color channels span across multiple cardiac cycles, highlighting the complexity of the problem, and providing guidance for PPG segment size selection (i.e., a PPG segment should consist of at least a couple of cardiac cycles). Finally, Fig. \ref{fig:shap} (f)-(h) depict some low-quality PPG segments, that contain some irregularities, e.g., motion artifacts, noise etc. Nevertheless, the SHAP method still identifies crucial features in these abnormal-looking PPG segments successfully. This illustrates the robustness of the LSTM-FCN model, and the efficacy of the SHAP method in identifying the right set of features that influence the model's decisions.

%Subfigures (a) to (h) display examples of PPG waveforms over a duration of 3.3 seconds, corresponding to an input array size of 100x3 - the standard input for the model. 

%% file: conclusion.tex
\section{Conclusion}
\label{sec:conclusion} 

We utilized the camera of a regular smartphone to record a small video of the fingertip for non-invasive dehydration monitoring, exploiting the fact that the PPG signal extracted from the video data contains the variations in the peripheral blood volume due to change in hydration level of a person over time. To train and test our AI models, we constructed a custom multi-session labeled dataset by collecting video-PPG data from 25 subjects who were fasting during Ramadan in 2023. With this, we solved two distinct problems: 1) dehydration detection, 2) dehydration level classification on a scale of 1 to 4. For both classification problems, our ML, DL and transformer models provided a very high accuracy, i.e., mostly in the range of 95\% to 99\%. We also proposed a highly-accurate alternate method where we fed the high-dimensional PPG time-series data to a DL model for feature extraction, followed by t-SNE method for feature selection and dimensionality reduction, followed by a number of ML classifiers that did dehydration level classification. Finally, we interpreted the decisions by our DL model under the SHAP-based explainable AI framework. 

The proposed smartphone-based non-invasive method for dehydration monitoring is highly accurate, allows frequent do-it-yourself monitoring, is easy to use, and provides rapid results. Further, the proposed method is cost-effective and thus inline with the sustainable development goals 3 \& 10 of the United Nations, and a step-forward to patient-centric healthcare systems, smart homes, and smart cities of future. The anticipated beneficiaries of the proposed method include: sportsmen, athletes, elderly, diabetic and diarrhea patients, and labor people working outdoors.

%% file: main.bbl
\begin{thebibliography}{10}
\providecommand{\url}[1]{#1}
\csname url@rmstyle\endcsname
\providecommand{\newblock}{\relax}
\providecommand{\bibinfo}[2]{#2}
\providecommand\BIBentrySTDinterwordspacing{\spaceskip=0pt\relax}
\providecommand\BIBentryALTinterwordstretchfactor{4}
\providecommand\BIBentryALTinterwordspacing{\spaceskip=\fontdimen2\font plus
\BIBentryALTinterwordstretchfactor\fontdimen3\font minus
  \fontdimen4\font\relax}
\providecommand\BIBforeignlanguage[2]{{%
\expandafter\ifx\csname l@#1\endcsname\relax
\typeout{** WARNING: IEEEtran.bst: No hyphenation pattern has been}%
\typeout{** loaded for the language `#1'. Using the pattern for}%
\typeout{** the default language instead.}%
\else
\language=\csname l@#1\endcsname
\fi
#2}}

\bibitem{1d}
A.~Breland, Ed., \emph{\BIBforeignlanguage{en}{Handbook of pediatric
  surgery}}.\hskip 1em plus 0.5em minus 0.4em\relax USA: American Medical,
  Sept. 2023.

\bibitem{2d}
E.~J{\'e}quier and F.~Constant, ``\BIBforeignlanguage{en}{Water as an essential
  nutrient: the physiological basis of hydration},''
  \emph{\BIBforeignlanguage{en}{Eur. J. Clin. Nutr.}}, vol.~64, no.~2, pp.
  115--123, Feb. 2010.

\bibitem{Cooper2000-xq}
G.~Cooper, \emph{The cell: A molecular approach}, 2nd~ed.\hskip 1em plus 0.5em
  minus 0.4em\relax Sunderland, MA: Sinauer Associates, Aug. 2000.

\bibitem{Thornton2016-nt}
S.~N. Thornton, ``\BIBforeignlanguage{en}{Increased hydration can be associated
  with weight loss},'' \emph{\BIBforeignlanguage{en}{Front. Nutr.}}, vol.~3,
  p.~18, June 2016.

\bibitem{Sahoo2014-fn}
S.~Sahoo, M.~K. Aurich, J.~J. Jonsson, and I.~Thiele,
  ``\BIBforeignlanguage{en}{Membrane transporters in a human genome-scale
  metabolic knowledgebase and their implications for disease},''
  \emph{\BIBforeignlanguage{en}{Front. Physiol.}}, vol.~5, p.~91, Mar. 2014.

\bibitem{Lee2015-pl}
D.~Lee, W.~Hwang, M.~Artan, D.-E. Jeong, and S.-J. Lee,
  ``\BIBforeignlanguage{en}{Effects of nutritional components on aging},''
  \emph{\BIBforeignlanguage{en}{Aging Cell}}, vol.~14, no.~1, pp. 8--16, Feb.
  2015.

\bibitem{Genuis2013-ai}
S.~J. Genuis, M.~E. Sears, G.~Schwalfenberg, J.~Hope, and R.~Bernhoft,
  ``\BIBforeignlanguage{en}{Clinical detoxification: elimination of persistent
  toxicants from the human body},''
  \emph{\BIBforeignlanguage{en}{ScientificWorldJournal}}, vol. 2013, p. 238347,
  June 2013.

\bibitem{Genuis2011-ux}
S.~J. Genuis, D.~Birkholz, I.~Rodushkin, and S.~Beesoon,
  ``\BIBforeignlanguage{en}{Blood, urine, and sweat ({BUS}) study: monitoring
  and elimination of bioaccumulated toxic elements},''
  \emph{\BIBforeignlanguage{en}{Arch. Environ. Contam. Toxicol.}}, vol.~61,
  no.~2, pp. 344--357, Aug. 2011.

\bibitem{Keep2012-hc}
R.~F. Keep, Y.~Hua, and G.~Xi, ``\BIBforeignlanguage{en}{Brain water content. a
  misunderstood measurement?}'' \emph{\BIBforeignlanguage{en}{Transl. Stroke
  Res.}}, vol.~3, no.~2, pp. 263--265, June 2012.

\bibitem{EFSA_Panel}
{EFSA Panel on Dietetic Products, Nutrition, and Allergies (NDA)}, ``Scientific
  opinion on dietary reference values for water,'' \emph{EFSA J.}, vol.~8,
  no.~3, Mar. 2010.

\bibitem{12d}
D.~R. Thomas, T.~R. Cote, L.~Lawhorne, S.~A. Levenson, L.~Z. Rubenstein, D.~A.
  Smith, R.~G. Stefanacci, and E.~G. Tangalos, ``Morley and dehydration
  council, ``understanding clinical dehydration and its treatment,''
  \emph{Journal of the American Medical Directors Association}, vol.~9, no.~5,
  pp. 292--301, 2008.

\bibitem{13d}
A.~M. El-Sharkawy, A.~Virdee, A.~Wahab, D.~J. Humes, O.~Sahota, M.~A.~J.
  Devonald, and D.~N. Lobo, ``Dehydration and clinical outcome in hospitalised
  older adults: A cohort study,'' \emph{Eur. Geriatr. Med.}, vol.~8, no.~1, pp.
  22--29, Feb. 2017.

\bibitem{16d}
H.~M. Logan-Sprenger, G.~J.~F. Heigenhauser, G.~L. Jones, and L.~L. Spriet,
  ``\BIBforeignlanguage{en}{The effect of dehydration on muscle metabolism and
  time trial performance during prolonged cycling in males},''
  \emph{\BIBforeignlanguage{en}{Physiol. Rep.}}, vol.~3, no.~8, p. e12483, Aug.
  2015.

\bibitem{14d}
A.~M. El-Sharkawy, O.~Sahota, R.~J. Maughan, and D.~N. Lobo,
  ``\BIBforeignlanguage{en}{The pathophysiology of fluid and electrolyte
  balance in the older adult surgical patient},''
  \emph{\BIBforeignlanguage{en}{Clin. Nutr.}}, vol.~33, no.~1, pp. 6--13, Feb.
  2014.

\bibitem{15d}
E.~Dupouy and M.~Gurinovic, ``\BIBforeignlanguage{en}{Sustainable food systems
  for healthy diets in europe and central asia: Introduction to the special
  issue},'' \emph{\BIBforeignlanguage{en}{Food Policy}}, vol.~96, no. 101952,
  p. 101952, Oct. 2020.

\bibitem{m1}
L.~E. Armstrong, ``Assessing hydration status: the elusive gold standard,''
  \emph{Journal of the American College of Nutrition}, vol.~26, no. sup5, pp.
  575S--584S, 2007.

\bibitem{m2}
------, ``Hydration assessment techniques,'' \emph{Nutrition reviews}, vol.~63,
  no. suppl\_1, pp. S40--S54, 2005.

\bibitem{m3}
K.~J. Sollanek, R.~W. Kenefick, S.~N. Cheuvront, and R.~S. Axtell, ``Potential
  impact of a 500-ml water bolus and body mass on plasma osmolality dilution,''
  \emph{European journal of applied physiology}, vol. 111, pp. 1999--2004,
  2011.

\bibitem{m4}
\BIBentryALTinterwordspacing
L.~E. Armstrong, J.~A. Herrera~Soto, F.~T. Hacker, D.~J. Casa, S.~A. Kavouras,
  and C.~M. Maresh, ``Urinary indices during dehydration, exercise, and
  rehydration,'' \emph{Intl. Journal of Sport Nutrition}, vol.~8, no.~4, pp.
  345--355, Dec. 1998. [Online]. Available:
  \url{http://dx.doi.org/10.1123/ijsn.8.4.345}
\BIBentrySTDinterwordspacing

\bibitem{m5}
D.~L. Smith, I.~Shalmiyeva, J.~Deblois, and M.~Winke,
  ``\BIBforeignlanguage{en}{Use of salivary osmolality to assess
  dehydration},'' \emph{\BIBforeignlanguage{en}{Prehosp. Emerg. Care}},
  vol.~16, no.~1, pp. 128--135, Jan. 2012.

\bibitem{m6}
D.~C. Knottenbelt, ``The urinary system,'' in \emph{The Equine Manual}.\hskip
  1em plus 0.5em minus 0.4em\relax Elsevier, 2006, pp. 659--712.

\bibitem{carrieri2020explainable}
A.~P. Carrieri, N.~Haiminen, S.~Maudsley-Barton, L.-J. Gardiner, B.~Murphy,
  A.~Mayes, S.~Paterson, S.~Grimshaw, M.~Winn, C.~Shand, \emph{et~al.},
  ``Explainable ai reveals key changes in skin microbiome associated with
  menopause, smoking, aging and skin hydration,'' \emph{bioRxiv}, pp. 2020--07,
  2020.

\bibitem{ahsannaturepaper}
A.~Mehmood, A.~Sarouji, M.~M.~U. Rahman, and T.~Y. Al-Naffouri, ``Your
  smartphone could act as a pulse-oximeter and as a single-lead ecg,''
  \emph{Scientific Reports}, vol.~13, no.~1, p. 19277, 2023.

\bibitem{liaqat2020non}
S.~Liaqat, K.~Dashtipour, K.~Arshad, and N.~Ramzan, ``Non invasive skin
  hydration level detection using machine learning,'' \emph{Electronics},
  vol.~9, no.~7, p. 1086, 2020.

\bibitem{liaqat2022personalized}
S.~Liaqat, K.~Dashtipour, A.~Rizwan, M.~Usman, S.~A. Shah, K.~Arshad,
  K.~Assaleh, and N.~Ramzan, ``Personalized wearable electrodermal
  sensing-based human skin hydration level detection for sports, health and
  wellbeing,'' \emph{Scientific Reports}, vol.~12, no.~1, p. 3715, 2022.

\bibitem{rizwan2020non}
A.~Rizwan, N.~A. Ali, A.~Zoha, M.~Ozturk, A.~Alomainy, M.~A. Imran, and Q.~H.
  Abbasi, ``Non-invasive hydration level estimation in human body using
  galvanic skin response,'' \emph{IEEE Sensors Journal}, vol.~20, no.~9, pp.
  4891--4900, 2020.

\bibitem{kulkarni2021non}
N.~Kulkarni, C.~Compton, J.~Luna, and M.~A.~U. Alam, ``A non-invasive
  context-aware dehydration alert system,'' in \emph{Proceedings of the 22nd
  International Workshop on Mobile Computing Systems and Applications}, 2021,
  pp. 157--159.

\bibitem{S2023SmartphoneBS}
P.~Adith, M.~Guhan, R.~Praveen~Kumar, M.~Sri~Hari, and K.~Nalinadevi,
  ``Smartphone based skin hydration level sensor and sunburn prediction,'' in
  \emph{2023 14th International Conference on Computing Communication and
  Networking Technologies (ICCCNT)}, 2023, pp. 1--7.

\bibitem{reljin2018automatic}
N.~Reljin, Y.~Malyuta, G.~Zimmer, Y.~Mendelson, D.~J. Blehar, C.~E. Darling,
  and K.~H. Chon, ``Automatic detection of dehydration using support vector
  machines,'' in \emph{2018 14th Symposium on Neural Networks and Applications
  (NEUREL)}.\hskip 1em plus 0.5em minus 0.4em\relax IEEE, 2018, pp. 1--6.

\bibitem{posada2019mild}
H.~F. Posada-Quintero, N.~Reljin, A.~Moutran, D.~Georgopalis, E.~C.-H. Lee,
  G.~E. Giersch, D.~J. Casa, and K.~H. Chon, ``Mild dehydration identification
  using machine learning to assess autonomic responses to cognitive stress,''
  \emph{Nutrients}, vol.~12, no.~1, p.~42, 2019.

\bibitem{suryadevara2015towards}
N.~K. Suryadevara, S.~C. Mukhopadhyay, and L.~Barrack, ``Towards a smart
  non-invasive fluid loss measurement system,'' \emph{Journal of medical
  systems}, vol.~39, pp. 1--10, 2015.

\bibitem{Sabry2022}
\BIBentryALTinterwordspacing
F.~Sabry, T.~Eltaras, W.~Labda, F.~Hamza, K.~Alzoubi, and Q.~Malluhi, ``Towards
  on-device dehydration monitoring using machine learning from wearable
  device's data,'' \emph{Sensors}, vol.~22, no.~5, p. 1887, Feb. 2022.
  [Online]. Available: \url{http://dx.doi.org/10.3390/s22051887}
\BIBentrySTDinterwordspacing

\bibitem{alvarez2019machine}
A.~Alvarez, E.~Severeyn, J.~Velasquez, S.~Wong, G.~Perpinan, and M.~Huerta,
  ``Machine learning methods in the classification of the athletes
  dehydration,'' in \emph{2019 IEEE Fourth Ecuador Technical Chapters Meeting
  (ETCM)}.\hskip 1em plus 0.5em minus 0.4em\relax IEEE, 2019, pp. 1--5.

\bibitem{7320006}
M.~Ring, C.~Lohmueller, M.~Rauh, and B.~M. Eskofier, ``On sweat analysis for
  quantitative estimation of dehydration during physical exercise,'' in
  \emph{2015 37th Annual Intl. Conf. of the IEEE Engg. in Medicine and Biology
  Society (EMBC)}, 2015, pp. 7011--7014.

\bibitem{7182268}
M.~Ring, C.~Lohmueller, M.~Rauh, J.~Mester, and B.~M. Eskofier, ``A
  temperature-based bioimpedance correction for water loss estimation during
  sports,'' \emph{IEEE Journal of Biomedical and Health Informatics}, vol.~20,
  no.~6, pp. 1477--1484, 2016.

\bibitem{7539317}
M.~Ring, C.~Lohmueller, M.~Rauh, J.~Mester, and B.~Eskofier, ``Salivary markers
  for quantitative dehydration estimation during physical exercise,''
  \emph{IEEE Journal of Biomedical and Health Informatics}, vol.~21, no.~5, pp.
  1306--1314, 2017.

\bibitem{7759295}
Y.~Mengistu, M.~Pham, H.~Manh~Do, and W.~Sheng, ``Autohydrate: A wearable
  hydration monitoring system,'' in \emph{2016 IEEE/RSJ Intl. Conf. on
  Intelligent Robots and Systems (IROS)}, 2016, pp. 1857--1862.

\bibitem{sm13}
P.~Saha, S.~M.~I. Zulfiker, T.~Hashem, and K.~A. Islam, ``Dehydration scan: An
  artificial intelligence assisted smartphone-based system for early detection
  of dehydration,'' in \emph{Lecture Notes of the Institute for Computer
  Sciences, Social Informatics and Telecommunications Engineering}, ser.
  Lecture notes of the Institute for Computer Sciences, Social Informatics and
  Telecommunications Engineering.\hskip 1em plus 0.5em minus 0.4em\relax Cham:
  Springer Nature Switzerland, 2023, pp. 289--312.

\bibitem{sm14}
C.~Liu, F.~Tsow, D.~Shao, Y.~Yang, R.~Iriya, and N.~Tao, ``Skin mechanical
  properties and hydration measured with mobile phone camera,'' \emph{IEEE
  Sensors Journal}, vol.~16, no.~4, pp. 924--930, 2015.

\bibitem{hasanhydrationpaper}
H.~M. Buttar, K.~Pervez, M.~Rahman, K.~Riaz, and Q.~H. Abbasi, ``Non-contact
  monitoring of dehydration using rf data collected off the chest and the
  hand,'' \emph{IEEE Sensors journal}, 2023.

\bibitem{s22051887}
\BIBentryALTinterwordspacing
F.~Sabry, T.~Eltaras, W.~Labda, F.~Hamza, K.~Alzoubi, and Q.~Malluhi, ``Towards
  on-device dehydration monitoring using machine learning from wearable
  device's data,'' \emph{Sensors}, vol.~22, no.~5, 2022. [Online]. Available:
  \url{https://www.mdpi.com/1424-8220/22/5/1887}
\BIBentrySTDinterwordspacing

\bibitem{Ring2017-ua}
M.~Ring, C.~Lohmueller, M.~Rauh, J.~Mester, and B.~M. Eskofier, ``Quantitative
  dehydration estimation,'' 2017.

\bibitem{Kirby2021-jn}
M.~Kirby, ``{SpectroPhon} {DBM} subject data,'' 2021.

\bibitem{rabiasensorsletterspaper}
R.~Ahmed, A.~Mehmood, M.~M.~U. Rahman, and O.~A. Dobre, ``A deep learning \&
  fast wavelet transform-based hybrid approach for denoising of ppg signals,''
  \emph{IEEE Sensors Letters}, 2023.

\bibitem{NIPS2017_3f5ee243}
A.~Vaswani, N.~Shazeer, N.~Parmar, J.~Uszkoreit, L.~Jones, A.~N. Gomez, L.~u.
  Kaiser, and I.~Polosukhin, ``Attention is all you need,'' in \emph{Advances
  in Neural Information Processing Systems}, I.~Guyon, U.~V. Luxburg,
  S.~Bengio, H.~Wallach, R.~Fergus, S.~Vishwanathan, and R.~Garnett, Eds.,
  vol.~30.\hskip 1em plus 0.5em minus 0.4em\relax Curran Associates, Inc.,
  2017.

\bibitem{Song_Rajan_Thiagarajan_Spanias_2018}
\BIBentryALTinterwordspacing
H.~Song, D.~Rajan, J.~Thiagarajan, and A.~Spanias, ``Attend and diagnose:
  Clinical time series analysis using attention models,'' \emph{Proceedings of
  the AAAI Conference on Artificial Intelligence}, vol.~32, no.~1, Apr. 2018.
  [Online]. Available:
  \url{https://ojs.aaai.org/index.php/AAAI/article/view/11635}
\BIBentrySTDinterwordspacing

\bibitem{8983326}
G.~Yan, S.~Liang, Y.~Zhang, and F.~Liu, ``Fusing transformer model with
  temporal features for ecg heartbeat classification,'' in \emph{2019 IEEE
  Intl. Conf. on Bioinformatics and Biomedicine (BIBM)}, 2019, pp. 898--905.

\bibitem{ZEYNALI2023105130}
\BIBentryALTinterwordspacing
M.~Zeynali, H.~Seyedarabi, and R.~Afrouzian, ``Classification of eeg signals
  using transformer based deep learning and ensemble models,'' \emph{Biomedical
  Signal Processing and Control}, vol.~86, p. 105130, 2023. [Online].
  Available:
  \url{https://www.sciencedirect.com/science/article/pii/S1746809423005633}
\BIBentrySTDinterwordspacing

\bibitem{tahir2024cuffless}
M.~A. Tahir, A.~Mehmood, M.~M.~U. Rahman, M.~W. Nawaz, K.~Riaz, and Q.~H.
  Abbasi, ``Cuff-less arterial blood pressure waveform synthesis from
  single-site ppg using transformer \& frequency-domain learning,'' \emph{Arxiv
  preprint arXiv2401.05452}, 2024.

\bibitem{lundberg2017unified}
S.~Lundberg and S.-I. Lee, ``A unified approach to interpreting model
  predictions,'' 2017.

\bibitem{shapxai}
\BIBentryALTinterwordspacing
M.~Sundararajan, A.~Taly, and Q.~Yan, ``Axiomatic attribution for deep
  networks,'' in \emph{Int. Conf. on Machine Learning}, 2017. [Online].
  Available: \url{https://api.semanticscholar.org/CorpusID:16747630}
\BIBentrySTDinterwordspacing

\end{thebibliography}
